\documentclass{article}

\PassOptionsToPackage{numbers, compress}{natbib}
\usepackage[preprint]{neurips_2026}          

\usepackage[utf8]{inputenc}
\usepackage[T1]{fontenc}
\usepackage{hyperref}
\usepackage{url}
\usepackage{booktabs}
\usepackage{makecell}
\usepackage{multirow}
\usepackage{caption}
\usepackage{amsfonts}
\usepackage{amsmath}
\usepackage{amssymb}
\usepackage{nicefrac}
\usepackage{microtype}
\usepackage{threeparttable} 
\usepackage{graphicx}
\usepackage{dsfont}
\usepackage{tabularx}       
\usepackage{enumitem}        

\usepackage{pifont}          
\usepackage{algorithm}       
\usepackage{algpseudocode}   
\usepackage{pgfplots}        
\pgfplotsset{compat=1.18}
\newcommand{\cmark}{\ding{51}}

\usepackage{tcolorbox}
\tcbuselibrary{skins,breakable}

\definecolor{chatBlue}{HTML}{1565C0}
\definecolor{chatGreen}{HTML}{2E7D32}
\definecolor{chatOrange}{HTML}{E65100}
\definecolor{chatRed}{HTML}{C62828}
\definecolor{chatPurple}{HTML}{6A1B9A}
\definecolor{chatTeal}{HTML}{00838F}

\newtcolorbox{chatbox}[1][Advisor Note]{%
  colback=chatBlue!5!white,
  colframe=chatBlue!70!black,
  fonttitle=\sffamily\bfseries,
  title={#1},
  breakable,
  left=4pt, right=4pt, top=4pt, bottom=4pt,
}

\graphicspath{{figures/}}

\title{SceneFactory: GPU-Accelerated Multi-Agent Driving Simulation with Physics-Based Vehicle Dynamics}

\author{%
  Yicheng Zhu \\
  Rochester Institute of Technology\\
  \texttt{yz8733@rit.edu} \\
  \And
  Yang Chen \\
  Rochester Institute of Technology\\
  \texttt{yz8786@rit.edu} \\
  \And
  Tao Li \\
  City University of Hong Kong\\
  \texttt{li.tao@cityu.edu.hk} \\
  \And
  Zilin Bian \thanks{Corresponding author.}\\
  Rochester Institute of Technology\\
  \texttt{zxbite@rit.edu} \\
}

\begin{document}

\maketitle

\begin{figure}[H]
    \centering
    \includegraphics[width=0.8\linewidth]{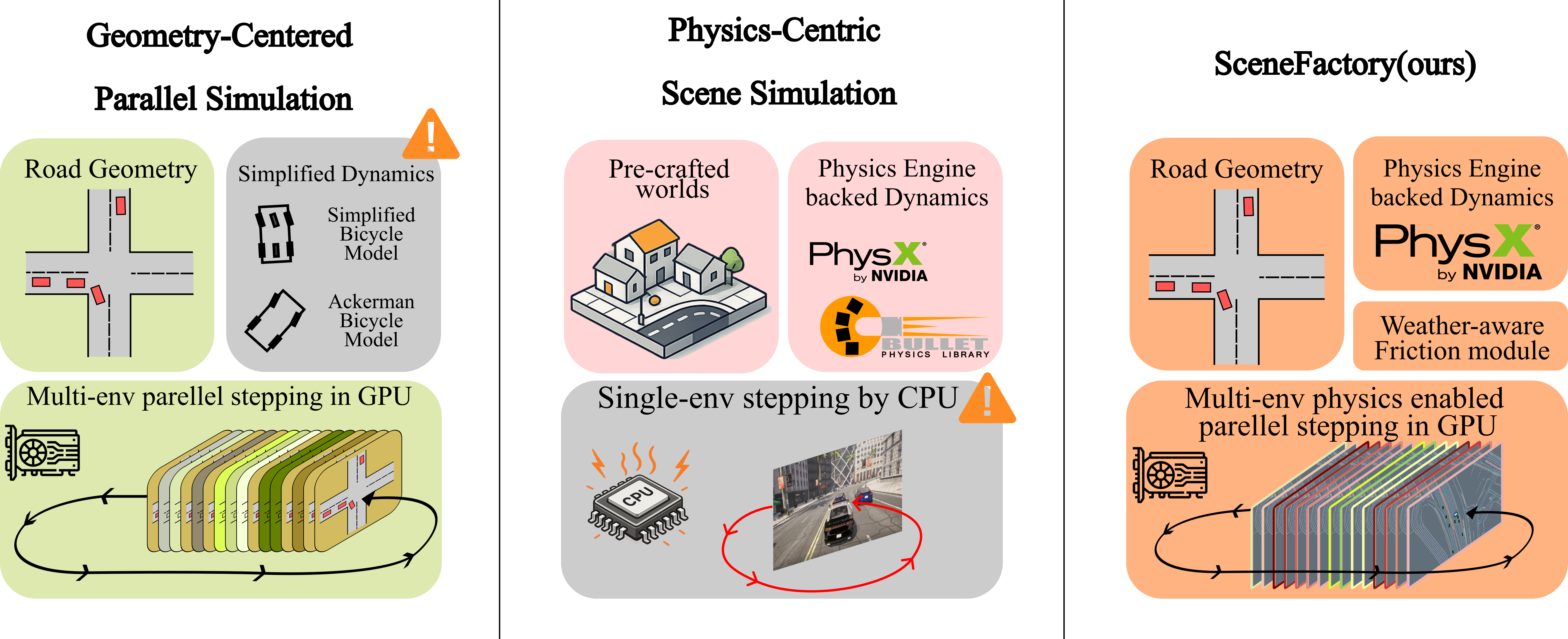}
    \caption{Existing autonomous-driving simulators typically choose between scalable geometry-centered batching and physics-centric scene simulation.
GPU-batched simulators execute many road layouts in parallel but use simplified vehicle dynamics, while physics-based simulators provide articulated rigid-body dynamics but lack native GPU-vectorized multi-world execution.
SceneFactory combines dataset-derived road geometry, PhysX articulated vehicles, weather-aware friction modeling, and tensor-vectorized GPU stepping to enable parallel multi-agent simulation across many independent driving scenes.}
    \label{fig:placeholder}
\end{figure}

\begin{abstract}
Autonomous-driving simulators typically trade physical fidelity for scalable parallelism. 
Physics-based platforms such as CARLA and MetaDrive provide articulated vehicle dynamics and contact, but their non-vectorized control interfaces make large batched training difficult. 
GPU-batched systems such as Waymax and GPUDrive scale to hundreds of scenarios by replacing rigid-body physics with simplified kinematics models, omitting tire--road interaction, suspension, contact dynamics, and road-condition-dependent friction. 
We introduce \textbf{SceneFactory}, a GPU-vectorized platform for procedural scene construction, physics-based multi-agent simulation, and reinforcement learning in autonomous driving environments. 
Built on NVIDIA Isaac Sim and Isaac Lab, SceneFactory represents worlds and agents as batched tensors: vehicle control, observations, rewards, resets, and policy inference are executed as GPU tensor operations over the Isaac Lab tensor API. 
SceneFactory converts Waymo Open Motion Dataset road topologies into simulation-ready USD (Universal Scene Description) worlds. SceneFactory runs many worlds concurrently on one GPU, populates each with multiple articulated PhysX vehicles, and maps precipitation and road-surface type to PhysX material friction coefficients. 
Thanks to the GPU vectorization, SceneFactory achieves up to \textbf{127$\times$} higher throughput than a non-vectorized PhysX baseline on the same GPU and physics solver, reaching \textbf{19{,}250 controlled-agent simulation steps per second (CASPS)} at 256 worlds $\times$ 16 agents. 
Cross-simulator transfer reveals an asymmetric dynamics gap: physics-grounded RL driving policies transfer to a simplified kinematic bicycle model with 99.5\% success, whereas the reverse transfer success rate drops to 47.3\%. 
Under wet-road friction, friction-aware policies reduce mean peak deceleration rate to avoid crash (DRAC) from 58.7 to 27.8\,m/s$^2$ without sacrificing goal reach. 
SceneFactory shows that scalable autonomous-driving training need not discard articulated rigid-body dynamics or physically grounded road-condition variation. Code will be released at \href{https://github.com/BrainCrackLab/SceneFactory}{GitHub}.
\end{abstract}

\section{Introduction}
\label{sec:introduction}

Simulation is central to autonomous-driving research, enabling closed-loop policy learning, controlled evaluation, and rare-event testing beyond what can be safely collected on public roads. 
However, simulators for learning-based autonomous driving must satisfy two requirements that are rarely met together. 
They must be scalable enough to generate experience across many independent scenes and agents, while remaining physically grounded enough to expose policies to the vehicle dynamics that determine safety-critical behavior. 
This physical fidelity is not merely a matter of visual realism. 
Tire--road interaction, braking, cornering, contact, suspension, and road-surface friction directly shape whether a control policy remains safe under closed-loop execution. 
Various weather situations further amplify this requirement: reduced pavement friction lengthens stopping distances, narrows the feasible control envelope, and can induce failures that are absent under nominal dry-road assumptions. 
Traffic safety statistics indicate that many weather-related crashes involve pavement-friction changes~\cite{fhwa_how_nodate}. 
Thus, a simulator that ignores rigid-body dynamics and road-condition-dependent friction may miss precisely the failure modes that physically grounded autonomous-driving evaluation should reveal.

Existing simulators typically occupy only one side of this fidelity--scale trade-off. 
Physics-based platforms such as CARLA~\cite{Dosovitskiy17} and MetaDrive~\cite{li2022metadrive} provide rich geometry, collision handling, and articulated vehicle dynamics through general-purpose simulation engines. 
Yet their control and environment interfaces are not designed as GPU-resident batched tensor programs over many independent worlds and agents. 
Vehicle state queries, action application, observation construction, and reward logic often pass through per-agent or per-scene CPU/Python calls. 
As the number of vehicles and scenarios grows, this non-vectorized path becomes a throughput bottleneck; large-scale evaluation typically relies on sequential rollouts, multiple simulator instances, or substantial interface engineering. 
GPU-accelerated systems such as Waymax~\cite{gulino_waymax_2023} and GPUDrive~\cite{kazemkhani_gpudrive_2025} take the opposite design point: they batch hundreds of scenarios as tensor computations on a single accelerator, but replace rigid-body physics with kinematic or bicycle dynamics. 
This abstraction enables scale, but omits the tire forces, suspension, contact dynamics, and road-surface-dependent friction needed for physically grounded evaluation. 
The missing capability is therefore \emph{batched physical simulation}: many independent driving scenes, each with multiple closed-loop agents, advanced in parallel while retaining physics-engine vehicle dynamics.


To fill this gap, we present \textbf{SceneFactory}, a GPU-accelerated, physically grounded, and weather-aware platform for multi-agent autonomous-driving simulation and training. 
Built on NVIDIA Isaac Sim and Isaac Lab~\cite{nvidia_isaac_2025}, SceneFactory uses PhysX to simulate articulated vehicles while exposing worlds and agents as batched tensors. 
Its core technical contribution is a \emph{tensor-vectorized articulated vehicle interface}: joint-state queries, actuation, observation assembly, reward computation, reset logic, and policy inference are expressed as GPU tensor operations over the Isaac Lab tensor API, eliminating per-agent Python control loops. 
SceneFactory converts Waymo Open Motion Dataset road topologies~\cite{Kan_2024_icra} into simulation-ready USD environments with lane geometry, road boundaries, and dataset-derived origin--destination tasks, and runs many such worlds concurrently on one GPU. 
To support road-condition-aware evaluation, SceneFactory further includes a weather-to-friction parameterization~\cite{zhao_tire-pavement_2024} that maps precipitation and road-surface type to PhysX material friction coefficients, allowing the same driving scene to be replayed under different physical conditions.

Our contributions are: 
(i) a procedural pipeline converting Waymo scenarios into GPU-parallelized multi-world Isaac Sim stages with multi-agent articulated PhysX vehicles; 
(ii) a tensor-vectorized vehicle interface that batches control, state queries, observations, rewards, resets, and policy inference across worlds and agents; 
(iii) a weather-to-friction model for per-world PhysX material parameters; and 
(iv) empirical validation showing \textbf{19{,}250 controlled-agent simulation steps per second (CASPS)} at 256 worlds $\times$ 16 agents, a \textbf{127$\times$} speedup over a non-vectorized PhysX baseline, and an \textbf{80.1\%} goal-reach rate for PPO policies trained on 128 Waymo topologies and evaluated on 64 held-out layouts.

\section{Related Work}
\label{sec:related_work}
\subsection{Driving Simulators}
\label{sec:rw_simulators}

As training autonomous agents in the real world is expensive and unsafe, simulation is essential.
Open-source platforms span a wide range of physics fidelity, scene diversity, and computational scalability.
CARLA~\cite{Dosovitskiy17}, built on Unreal Engine~4 with PhysX vehicle physics, provides rich articulated-body dynamics and detailed sensor simulation, but is CPU-bound, single-scene, and requires labor-intensive handcrafted maps.
MetaDrive~\cite{li2022metadrive} improves scene diversity by importing layouts from datasets such as Waymo Open Motion~\cite{Kan_2024_icra} or via procedural generation, but lacks GPU acceleration, supports only one scene per process, and has no automatic weather-to-friction coupling.
GPUDrive~\cite{kazemkhani_gpudrive_2025}, built on the Madrona game engine~\cite{shacklett_madrona_2023}, achieves over a million steps per second for multi-scene, multi-agent training, but uses a kinematic bicycle model that cannot capture tire--road contact, suspension compliance, or friction-dependent dynamics.
Simplification of dynamics is a valid design choice for large-scale policy learning where map-following dominates; SceneFactory targets the complementary setting where physics fidelity and weather-aware friction directly affect evaluation validity---for example, safety-critical analysis under adverse road conditions.

\begin{table*}[t]
\centering
\footnotesize
\setlength{\tabcolsep}{3.5pt}
\begin{threeparttable}
\begin{tabular}{l cccccccc}
\toprule
\textbf{Simulator}
  & \rotatebox{75}{\textbf{Multi-agent}}
  & \rotatebox{75}{\textbf{Batched GPU envs}$^\ddagger$}
  & \rotatebox{75}{\textbf{Sensor Sim}}
  & \rotatebox{75}{\textbf{Expert Data}}
  & \rotatebox{75}{\textbf{Sim-agents}}
  & \rotatebox{75}{\textbf{Rigid-body physics}}
  & \rotatebox{75}{\textbf{Weather$\to$Friction}$^\dagger$}
  & \rotatebox{75}{\textbf{Routes / Goals}} \\
\midrule
CARLA \cite{Dosovitskiy17} &  &  & \cmark &  & \cmark & \cmark &  & Waypoints \\
SUMMIT \cite{cai_summit_2020} & $\checkmark$ ($\geq$400) &  & \cmark &  & \cmark & \cmark &  & - \\
SMARTS \cite{zhou_smarts_2021} & \cmark &  &  &  &  & \cmark &  & Waypoints \\
nuPlan \cite{caesar_nuplan_2022} &  &  & \cmark & \cmark & \cmark &  &  & Waypoints \\
Nocturne \cite{vinitsky_nocturne_2022} & $\checkmark$ ($\geq$128) &  & \cmark & \cmark & \cmark &  &  & Goal point \\
MetaDrive \cite{li2022metadrive} & \cmark &  & \cmark & \cmark & \cmark & \cmark &  & - \\
TorchDriveSim \cite{lavington_torchdriveenv_2024} & \cmark & \cmark &  &  &  &  &  & - \\
Waymax \cite{gulino_waymax_2023} & $\checkmark$ ($\geq$128) & \cmark &  & \cmark & \cmark &  &  & Waypoints \\
GPUDrive \cite{kazemkhani_gpudrive_2025} & $\checkmark$ ($\geq$128) & \cmark & \cmark & \cmark & \cmark &  &  & Goal point \\
\textbf{SceneFactory (ours)} & $\checkmark$ ($\geq$128) & \cmark & \cmark & \cmark & \cmark & \cmark & \cmark & Goal point \\
\bottomrule
\end{tabular}
\begin{tablenotes}\footnotesize
  \item[$^\dagger$] Physics-engine friction set from precipitation + surface type; not just manual friction or visual weather.
  \item[$^\ddagger$] Single-process GPU-resident batched stepping; CARLA/MetaDrive use multi-process parallelism only.
\end{tablenotes}
\caption{Comparison of driving simulators. SceneFactory is, to our knowledge, the first to combine batched GPU execution, multi-agent support, rigid-body physics, and weather-to-friction modeling.}
\label{tab:simulators}
\end{threeparttable}
\end{table*}

\subsection{Weather-Aware Driving Behavior}
\label{sec:rw_weather}

Adverse weather affects both driver behavior and vehicle dynamics in ways that matter for safety-critical evaluation.
Chen et al.~\cite{chen_influence_2019} showed that drivers' perceived risk increases with worsening weather during car-following; Ahmed and Ghasemzadeh~\cite{ahmed_impacts_2018} found that the probability of speed reduction increases by 23--29\% in rain using SHRP2 naturalistic data; and Peng et al.~\cite{peng_examining_2018} demonstrated significant changes in traffic parameters under fog and rain.
Most autonomous driving systems achieve strong performance under the implicit assumption of favorable weather~\cite{li_autonomous_2025}, and existing adverse-weather research focuses primarily on perception degradation under rain, fog, or snow~\cite{song_weather_2020, lei23cola, fursa_worsening_2021, rahmati_edge_2025}.
Comparatively few works consider how vehicle dynamics change: Aghdasian et al.~\cite{aghdasian_tackling_2025} trained lane-keeping policies under snowy CARLA conditions, but without a physics-grounded friction model.
On the tire--pavement side, Zhao et al.~\cite{zhao_tire-pavement_2024} developed a friction model accounting for pavement texture and water film thickness, providing a physics-based mapping from environmental conditions to effective friction coefficients.
SceneFactory integrates this model directly into PhysX, so agents experience and must adapt to weather-dependent dynamics rather than a behavioral proxy.

\subsection{GPU-Accelerated Simulation Frameworks}
\label{sec:rw_gpu_frameworks}

The need for large-scale RL training has driven GPU-accelerated simulation.
Madrona~\cite{shacklett_madrona_2023} is a batch entity-component-system engine enabling thousands of independent worlds on a single GPU; it underpins GPUDrive.
Brax~\cite{freeman_brax_2021} provides differentiable JAX-based rigid-body simulation, primarily for locomotion.
Isaac Sim and Isaac Lab~\cite{nvidia_isaac_2025, mittal_orbit_2023} take a different approach: Isaac Sim provides a full PhysX~5 runtime with OpenUSD scene management, and Isaac Lab (evolved from Orbit~\cite{mittal_orbit_2023}) adds a vectorized multi-agent RL layer (\texttt{DirectMARLEnv}) with Fabric-based GPU-resident state cloning.
The combination of using USD scene management, articulated PhysX~5 vehicle dynamics, and an on-device training loop makes Isaac Lab the natural substrate for SceneFactory, where physics fidelity (articulated vehicles with tire friction, suspension, and contact sensing) and throughput (hundreds of vehicles across parallel worlds) must coexist.

\section{Simulator Design}
\label{sec:simulator_design}

We build on NVIDIA Isaac Sim and Isaac Lab~\cite{nvidia_isaac_2025, mittal_orbit_2023} and adopt the $k$-nearest-neighbor road-point encoding and per-entity MLP with max-pool aggregation from GPUDrive~\cite{kazemkhani_gpudrive_2025}.
Our contributions are in \emph{simulator infrastructure}, not in RL algorithms or policy architecture: we contribute procedural USD scene construction from Waymo data, articulated PhysX vehicle dynamics with system identification, and a physics-based weather-to-friction module, each described below.
The full three-phase pipeline (offline preprocessing, GPU environment instantiation, on-device PPO training loop) is detailed in Algorithm~\ref{alg:pipeline} in the appendix.

\begin{figure}[H]
    \centering
    \includegraphics[width=0.9\linewidth]{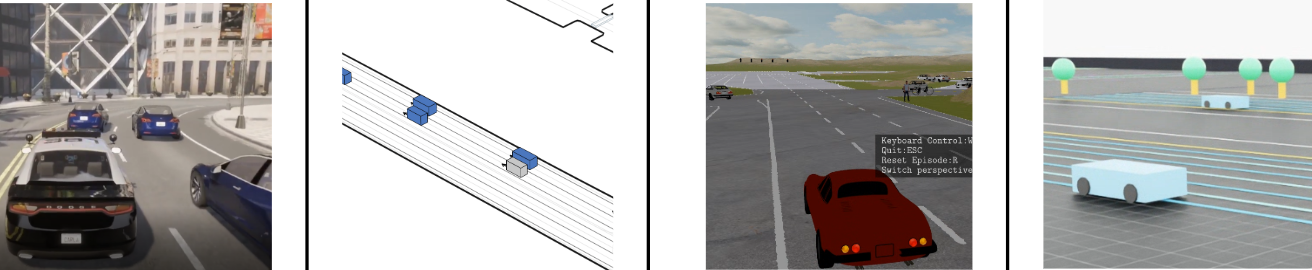}
    \caption{Qualitative comparison of four simulators. \textbf{CARLA} (1st from left) offers full rigid-body physics but renders a single scene sequentially. \textbf{GPUDrive} (2nd) parallelises thousands of scenes yet relies on a kinematic bicycle model. \textbf{MetaDrive} (3rd) supports rigid-body physics and procedural worlds but is limited to a single world at a time. \textbf{SceneFactory} (ours, right) unifies GPU-parallel multi-world execution with articulated PhysX vehicle dynamics.}
    \label{fig:comp_ss}
\end{figure}

\subsection{Data Processing and Procedural Scene Construction}
\label{sec:data_processing}

SceneFactory consumes driving scenarios from the Waymo Open Motion Dataset~\cite{Kan_2024_icra} and constructs physically simulated USD environments through an offline pipeline: an extraction script reads each TFRecord and produces one JSON per scenario containing road polylines (lane centers, road edges) and agent start/goal metadata.

\paragraph{Per-scenario world building.}
Each scene JSON is converted to a USD road environment by: re-centering all coordinates to the polyline centroid; z-flattening elevation to a flat ground plane; generating oriented box segments from consecutive polyline point pairs (gaps $>{3.0}$\,m and points outside the $400$\,m bounding box are skipped); and rendering segments as \texttt{PointInstancer} prims grouped by Waymo type.
Road geometry metadata is baked into USD \texttt{customData} and loaded into padded GPU tensors at startup, enabling batched lane-proximity queries with no USD access in the inner loop.
Segments are not PhysX collision objects; they serve as the geometric basis for rewards and observations.
Road prims are injected \emph{after} Isaac Lab's clone step so each environment hosts a distinct scenario (details in Appendix~\ref{app:scene_construction}).

Vehicle start/goal positions are filtered to the bounding box, capped at $M{=}16$ per world, and agents already within the 3.0\,m goal radius are discarded.

\subsection{Vehicle Asset Generation and System Identification}
\label{sec:vehicle_sysid}

SceneFactory uses a procedurally generated articulated vehicle rather than Isaac Sim's built-in PhysX Vehicle model.
The built-in model provides a validated tire model and drivetrain, but its internal state is managed outside the Fabric tensor pipeline, making it incompatible with batched GPU-resident cloning and preventing vectorized joint-state queries across the full agent population.
Our student vehicle is instead a 10-DOF articulated rigid body whose joint states, actuation commands, and contact materials are all accessible as GPU tensors; system identification against the built-in model recovers the dynamics fidelity lost by the simpler formulation.
The vehicle is defined by a parametric specification, converted to URDF, imported into Isaac Sim as a USD asset, and then calibrated via system identification.

\paragraph{Articulated vehicle structure.}
The vehicle is a 10-DOF articulated rigid body: a box chassis (1800\,kg, wheelbase 2.6\,m) with per-corner prismatic suspension joints, front revolute steering joints, and four wheel-spin joints.
The URDF is generated programmatically from a frozen parameter dataclass, imported into Isaac Sim, and post-processed to author PhysX contact materials and de-instance mesh prims for Fabric compatibility.
\paragraph{Action decoding.}
The policy outputs three continuous actions: throttle $\in [0,1]$, steering $\in [-1,1]$, and brake $\in [0,1]$.
Steering is decoded via a PD controller (Eq.~\ref{eq:steer_pd}); throttle drives the two front wheels; brake applies opposing torque to all four.
An external chassis wrench provides lateral damping ($-\lambda_{\mathrm{lat}} v_{\mathrm{lat}}$) and yaw damping ($-\lambda_{\mathrm{yaw}} \omega_z$) as calibrated approximations of slip-angle tire forces absent from the simplified contact solver. PD gains and torque limits are in Appendix~\ref{app:actions}.
\begin{equation}
\label{eq:steer_pd}
\tau_{\mathrm{steer}} = \mathrm{clamp}\!\bigl(K_p\,(\theta_{\max}\cdot\text{steering} - \theta) - K_d\,\dot{\theta},\;\pm\tau_{\mathrm{steer,max}}\bigr).
\end{equation}

\paragraph{System identification.}
The articulated vehicle model has 20 tunable dynamics parameters that are \emph{intrinsic} to the vehicle: drive/brake torques, steering PD gains, suspension stiffness/damping, wheel mass and inertia, chassis center-of-mass offset, and lateral/yaw stabilization coefficients.
These parameters determine how the vehicle responds to actuator commands; they do not change when the road surface changes.
Environment-level friction variation is handled separately by the PhysX contact solver via wheel material properties (\S\ref{sec:weather_friction}).

We fit these parameters via a five-stage cross-entropy method (CEM) against reference trajectories from the Isaac Sim built-in PhysX Vehicle model (the ``teacher''), which includes a validated tire model, drivetrain, and suspension.
The teacher executes 139 scripted maneuvers spanning longitudinal, lateral, combined, frequency-response, and surface-transition behaviors, covering the full operating envelope from 10\% to 100\% throttle.
The CEM evaluates 325 trials across five sequential stages (elite fraction 25\%), with stage-specific populations ranging from 49 to 97 samples; each stage inherits the best configuration from the prior stage.
Full loss formulation, maneuver breakdown, and per-stage results are in Appendix~\ref{app:sysid}.

\subsection{Weather-to-Friction Module}
\label{sec:weather_friction}

SceneFactory integrates a physics-based tire-pavement friction model that maps weather conditions to per-world friction coefficients, enabling training across a realistic range of road-surface conditions.

\paragraph{Friction model.}
We implement the modified Average Lumped LuGre (ALL) model of Zhao et al.~\cite{zhao_tire-pavement_2024}, which extends the LuGre bristle-deflection framework to account for pavement texture and water film.
Here $h$ denotes the water film thickness on the road surface (in mm), which the model uses together with a per-surface texture coefficient to compute the effective friction coefficient $\mu$ under partial hydroplaning.
The full formulation is provided in Appendix~\ref{app:friction_model}.

We provide three road surface presets calibrated to values in~\cite{zhao_tire-pavement_2024}:
\emph{Asphalt Concrete} (AC), \emph{Stone Mastic Asphalt} (SMA), and \emph{Open-Graded Friction Course} (OGFC).
Table~\ref{tab:friction_mu} (Appendix~\ref{app:friction_model}) reports $\mu$ values at representative water film depths.
At the reference speed of 13.89\,m/s, the ALL model predicts near-complete hydroplaning for $h \geq 1.0$\,mm on SMA surface; the simulator applies a floor of $\mu = 10^{-3}$ to avoid a degenerate zero-friction contact.
The ``hard rain'' condition used in the friction ablation (\S\ref{sec:ablation_friction}) corresponds to SMA at $h{=}2.0$\,mm, i.e.\ $\mu \approx 10^{-3}$.

\paragraph{Pipeline.}
Each world is assigned a road surface type and water film thickness $h_w$.
At construction time, the ALL model produces per-world static and dynamic friction coefficients at a reference speed of 13.89\,m/s.
Because all cloned environments share a single global ground plane (see \S\ref{sec:sharp_edges}), per-world friction variation is communicated to the policy through a 4-dimensional observation vector:
\begin{equation}
\label{eq:weather_obs}
\mathbf{o}_{\mathrm{weather}} = \bigl[\, h_w / h_{\mathrm{norm}},\; \mathds{1}_{\mathrm{AC}},\; \mathds{1}_{\mathrm{SMA}},\; \mathds{1}_{\mathrm{OGFC}} \,\bigr] \in \mathbb{R}^4,
\end{equation}
where $h_{\mathrm{norm}} = 1.0$\,mm and $\mathds{1}_{\{\cdot\}}$ is a one-hot surface type encoding.

\paragraph{RL interface.}
SceneFactory ships a ready-to-use RL interface with three configurable observation groups (ego state, road-geometry context, neighbor context), a continuous throttle/steer/brake action space shared across both the PhysX and bicycle-model backends (enabling zero-adapter policy transfer), and a ten-term reward function whose every weight is a YAML field.
The weather token $\boldsymbol{w}$ (Eq.~\ref{eq:weather_obs}) is injected directly into the ego observation, so a policy can condition on the current friction regime out of the box---a feature absent in simulators without physical weather support.
Full specifications are in Appendix~\ref{app:observations}--\ref{app:rewards}.

\paragraph{Usability and configuration.}
The entire platform is controlled through a single YAML hierarchy.
A researcher can:
(i)~\textbf{swap the scene pool} by pointing \texttt{scene\_factory.config\_path} at any preprocessed scene list;
(ii)~\textbf{scale parallelism} by changing \texttt{env.num\_envs} and \texttt{env.num\_agents\_per\_env}---no code changes required;
(iii)~\textbf{change the weather distribution} by editing surface-type probabilities and water-film ranges in the scene config YAML, which propagates through the ALL model and updates both PhysX materials and the observation token automatically;
(iv)~\textbf{switch dynamics backends} by setting \texttt{env.dynamics\_mode: bicycle} to replace PhysX with the lightweight kinematic integrator for rapid prototyping or transfer studies; and
(v)~\textbf{tune or ablate reward terms} by editing per-term weights without touching Python.
Training is launched with a single command, and evaluation replays a saved checkpoint against any scene pool with the same command and an \texttt{--eval} flag.
Compared to simulators such as CARLA~\cite{Dosovitskiy17}, which require separate Python APIs, scenario description files, and C++ recompilation to modify physics parameters, SceneFactory collapses the research iteration loop into config edits.

\subsection{Simulator Limitations and Sharp Edges}
\label{sec:sharp_edges}

SceneFactory makes several deliberate simplifications; a full catalogue with workarounds is in Appendix~\ref{app:sharp_edges}. All road geometry is z-flattened to $z{=}0$, causing bridges and overpasses to produce overlapping segments (mitigated by automated filters that reject scenes with excessive multi-level overlap). Waymo road-edge label noise (parking-lot entrances misclassified as road edges~\cite{kazemkhani_gpudrive_2025}) creates occasional goal conflicts that filtering reduces but cannot eliminate. Parked vehicles are excluded, removing static-obstacle interactions. Per-world friction is spatially uniform: the ALL-model $\mu$ is applied to each vehicle's wheel material at construction time, but all wheel contacts within a world share the same coefficient; local variation such as wet patches or surface transitions cannot be represented.

\section{Experiments}
\label{sec:experiments}

We evaluate SceneFactory across several dimensions:
simulation throughput (\S\ref{sec:throughput}),
end-to-end navigation performance on held-out Waymo scenes (\S\ref{sec:navigation_perf}),
a vehicle dynamics ablation quantifying the PhysX--bicycle physics gap (\S\ref{sec:ablation_dynamics}),
and a friction-awareness ablation (\S\ref{sec:ablation_friction}).
All experiments use a single shared navigation policy trained with PPO; the contribution of this work is the simulation platform itself, so we report policy performance as an end-to-end validation of the pipeline rather than as a claim of state-of-the-art driving.

\subsection{Experimental Setup}
\label{sec:exp_setup}

\paragraph{Hardware and software.}
All training and evaluation runs execute on a single NVIDIA RTX PRO 6000 Blackwell GPU (96\,GB VRAM).
The software stack consists of
NVIDIA Isaac Sim Full 5.1.0,
Isaac Lab 0.54.3 ~\cite{nvidia_isaac_2025},
RSL-RL 3.1.2 for PPO,
PyTorch 2.7.0+cu128, and
CUDA 12.8.
Physics runs at 120\,Hz (time step $\Delta t = 1/120$\,s);
the control policy acts at 30\,Hz (decimation factor 4).

\paragraph{Scene selection.}
From approximately 1{,}000 Waymo Open Motion Dataset scenarios~\cite{Kan_2024_icra},
we extract candidate road topologies through the pipeline described in \S\ref{sec:data_processing}.
Automated geometric filters reject scenes with degenerate lane structure, multi-level overlaps from z-flattening, or unreachable goals (see \S\ref{sec:sharp_edges}).
From the passing scenes we randomly sample 128 training scenes and a disjoint set of 64 held-out test scenes used throughout all main results; no manual inspection is performed.
Within each accepted scene, an automated filter then retains only those agents whose start and goal positions both lie within the 400\,m $\times$ 400\,m world bounding box, are separated by at least 3.0\,m, and have valid coordinates.
Up to 16 qualifying agents are extracted per scene; scenes with fewer valid agents leave the remaining agent slots inactive.

\paragraph{Training configuration.}
All runs use 256 parallel worlds $\times$ 16 agent slots; the number of unique road topologies $N$ varies by experiment.
Episodes last 50\,s; terminated agents are parked until the episode ends.
PPO: adaptive LR ($2.09{\times}10^{-4}$ initial, KL target 0.01), horizon 128, 5 epochs, 16 mini-batches, $\gamma{=}0.99$, $\lambda{=}0.98$, $\epsilon{=}0.15$ (Appendix~\ref{app:ppo_hparams}).

\paragraph{Evaluation protocol.}
We evaluate on 64 held-out test scenes (up to 16 agent slots each; 624 of 1{,}024 nominal slots contain a valid spawn, the rest being scenes with fewer than 16 qualifying moving vehicles).
The policy runs in deterministic mode under \emph{invincible mode}: events are logged but do not terminate agents, giving unconditional goal-reach rates.

We report four metrics: \textbf{success rate (SR)}, fraction of agents reaching their goal; \textbf{collision rate (CR)}, fraction triggering inter-vehicle contact above 25\,N; \textbf{DRAC} (deceleration rate to avoid crash), the minimum deceleration needed to avoid a neighbor from closing speed and gap ($v_{r}^2/2d$), reported as mean episode maximum DRAC ($>3.4$\,m/s$^2$, a standard highway threshold~\cite{fu_comparison_2021}). 

\subsection{Throughput}
\label{sec:throughput}

A core engineering contribution of SceneFactory is throughput: training with full rigid-body dynamics at a rate that makes large-scale PPO practical on a single GPU.
We compare SceneFactory against a non-vectorized predecessor built on the same Isaac Sim and PhysX~5 solver but using the built-in Vehicle Wizard model and Stable-Baselines3 (hereafter ``Baseline'').
Both pipelines use the same Waymo scene, 16 agents per world, identical observations, and the same RTX PRO 6000 GPU.
We report \emph{controlled agent steps per second} (CASPS): policy-controlled agents completing one 30\,Hz tick per wall-clock second, averaged over steady-state PPO iterations after a 50-iteration warmup.

We replicate a single Waymo scene at four scales (32--256 parallel worlds, 16 agents per world) and additionally sweep 1-agent-per-world configurations from 512 to 4{,}096 worlds.
Figure~\ref{fig:throughput_scaling} plots CASPS against total agent slots for all three series.
The Baseline stays flat at $\sim$152--174 CASPS regardless of world count, suggesting that per-agent Python loops (not the PhysX solver) are the bottleneck.
SceneFactory with 16 agents per world scales near-linearly, reaching \textbf{19{,}250 CASPS} at 256 worlds (4{,}096 slots), a \textbf{127$\times$} gain over the Baseline.
At 1 agent per world, inter-agent collision overhead is eliminated, yielding \textbf{41{,}742 CASPS} at 4{,}096 worlds, a further \textbf{2.2$\times$} over the dense configuration at equal agent-slot count.
Scaling to 8{,}192 worlds exceeds the 94.96\,GiB VRAM budget, establishing 4{,}096 worlds as the single-GPU ceiling on this hardware.

\begin{figure}[t]
\centering
\scalebox{0.82}{%
\begin{minipage}[c]{0.55\columnwidth}
  \begin{tikzpicture}
    \begin{axis}[
      width=\linewidth,
      height=4.5cm,
      xlabel={Agent slots},
      ylabel={CASPS},
      xmin=300, xmax=4700,
      xtick={512,1024,2048,4096},
      xticklabels={512,1{,}024,2{,}048,4{,}096},
      x tick label style={font=\scriptsize},
      ymode=log,
      ymin=100, ymax=60000,
      ytick={100,1000,10000},
      yticklabels={100,1{,}000,10{,}000},
      grid=major,
      grid style={dashed,gray!40},
      tick label style={font=\small},
      label style={font=\small},
      legend style={font=\tiny, at={(0.5,1.03)}, anchor=south, legend columns=3},
      mark size=2pt,
    ]
      \addplot[color=gray!70,mark=square*,thick,dashed] coordinates {
        (512,174)(1024,159)(2048,164)(4096,152)};
      \addlegendentry{Baseline}
      \addplot[color=blue!70,mark=*,thick] coordinates {
        (512,3870)(1024,7173)(2048,12225)(4096,19250)};
      \addlegendentry{SF, 16 ag/w}
      \addplot[color=orange!80!black,mark=triangle*,thick] coordinates {
        (512,14907)(1024,23397)(2048,32713)(4096,41742)};
      \addlegendentry{SF, 1 ag/w}
      \node[font=\scriptsize, blue!70, anchor=west] at (axis cs:4200,19250) {\textbf{127$\times$}};
      \node[font=\scriptsize, orange!80!black, anchor=west] at (axis cs:4200,41742) {\textbf{274$\times$}};
    \end{axis}
  \end{tikzpicture}
  \captionof{figure}{Throughput (CASPS, log scale). Blue: 16 ag/world; Orange: 1 ag/world; gray dashed: Baseline ($\sim$152--174).}
  \label{fig:throughput_scaling}
\end{minipage}%
\hspace{0.03\columnwidth}%
\begin{minipage}[c]{0.42\columnwidth}
  \centering\small
  \captionof{table}{Navigation generalization of the v8 policy (157M steps, invincible mode, 3.0\,m goal).}
  \label{tab:generalization}
  \vspace{2pt}
  \begin{tabular}{lccc}
  \toprule
  OD & Agents & SR $\uparrow$ & CR $\downarrow$ \\
  \midrule
  Original & 624     & 80.1\% & 4.6\% \\
  Random   & 1{,}024 & 87.0\% & 5.0\% \\
  \bottomrule
  \end{tabular}
\end{minipage}%
}
\end{figure}

MetaDrive (Bullet/CPU) peaks at $\sim$1{,}390 CASPS on the same machine; kinematic simulators (GPUDrive, Waymax) are not directly comparable due to their simpler per-step integration.
A full cross-simulator comparison with per-phase timing is in Appendix~\ref{app:throughput} (Table~\ref{tab:cross_simulator}).

\subsection{Navigation Performance}
\label{sec:navigation_perf}

We train using the v8 configuration (dry asphalt, no weather variation, 128 unique scenes across 256 worlds~$\times$~16 agents) and evaluate model\_300 ($\sim$157M environment steps) under two conditions that progressively decouple the test setting from training: (1)~64 held-out Waymo scenes with original dataset OD pairs, and (2)~the same scenes with destinations re-sampled at 10--100\,m travel distance.
All evaluations use invincible mode (see \S\ref{sec:exp_setup}) so that goal-reach rates are unconditional.

On held-out test scenes with original Waymo destinations, the policy achieves 80.1\% goal-reach (500 of 624 agents) after 157M steps.
With randomly sampled 10--100\,m destinations the rate rises to 87.0\% (891 of 1{,}024 agents), indicating that the original Waymo OD pairs include longer, more demanding routes.
For context, GPUDrive reports $>$98\% SR with a kinematic bicycle model; our vehicle dynamics ablation (\S\ref{sec:ablation_dynamics}) shows that a kinematic-bicycle-trained policy on the same scenes also achieves $\sim$100\% SR in its own environment, confirming that the lower PhysX SR reflects the additional difficulty of rigid-body dynamics rather than a policy or scene-selection artifact.

\subsection{Vehicle Dynamics Ablation}
\label{sec:ablation_dynamics}

To quantify the physics gap between the kinematic bicycle model and SceneFactory's articulated PhysX vehicle, we train two policies under otherwise identical conditions (same scenes, observations, rewards, and PPO hyperparameters) and cross-evaluate each in both simulators.
The bicycle environment is a drop-in replacement backend within SceneFactory: the same Isaac Lab RL loop, scene pool, observations, and rewards are used, but the PhysX articulated-body solver is replaced by a GPU-resident kinematic integrator that applies the standard front-wheel-drive bicycle equations at each 30\,Hz control tick.
It shares wheelbase (2.6\,m) and steering limit ($\pm$30\textdegree{}) with the PhysX vehicle but has no suspension, contact response, or lateral slip.

\begin{table}[t]
\centering
\caption{Physics-gap cross-evaluation on 64 held-out test scenes, 624 agents, invincible mode.
  ``Train env'' and ``Eval env'' denote the dynamics backend during training and evaluation, respectively.
  ``Bicycle'' is the GPU-batched kinematic model; ``PhysX'' is SceneFactory's 10-DOF articulated rigid body.}
\label{tab:dynamics_ablation}
\small
\begin{tabular}{llcc}
\toprule
Train env & Eval env & SR $\uparrow$ & CR $\downarrow$ \\
\midrule
Bicycle & Bicycle (in-distribution) & 100.0\% & 3.4\% \\
Bicycle & PhysX (transfer)          &  47.3\% & 14.7\% \\
PhysX   & Bicycle (transfer)        &  99.5\% & 48.2\% \\
PhysX   & PhysX (in-distribution)   &  80.1\% & 4.6\% \\
\bottomrule
\end{tabular}
\end{table}

The bicycle-trained policy achieves 100\% SR in-distribution but collapses to 47.3\% on PhysX, a 52.7\,pp drop driven by the absence of lateral slip and suspension compliance during training.
The PhysX policy transfers to the bicycle environment at 99.5\% SR, confirming richer dynamics generalize downward; the elevated 48.2\% CR reflects over-steering for forces absent in the kinematic model.

\subsection{Friction Conditioning}
\label{sec:ablation_friction}

We compare a friction-aware policy (v7: 10\% wet-world exposure, weather-context observation, $\sim$314M steps) against a friction-blind policy (v8: dry-only, $\sim$314M steps) on 64 held-out scenes.
Note that both checkpoints are trained longer than the v8 model reported in \S\ref{sec:navigation_perf} ($\sim$157M steps); the modest SR difference (80.1\% vs.\ 77.9\% on dry asphalt) reflects checkpoint-to-checkpoint variance, not a regression.

\begin{table}[h]
\centering
\caption{Friction conditioning ablation (64 test scenes, 624 agents, invincible mode). DRAC = deceleration rate to avoid crash (m/s$^2$).}
\label{tab:friction_ablation}
\small
\begin{tabular}{llccc}
\toprule
Surface & Policy & SR $\uparrow$ & CR $\downarrow$ & Mean max DRAC $\downarrow$ \\
\midrule
\multirow{2}{*}{\makecell[l]{Dry\\(AC, $\mu{=}1.1$)}}
  & v7(weather aware model) & 77.2\% & 5.1\% & 37.1 \\
  & v8(weather blind model) & 77.9\% & 6.3\% & 50.2 \\
\midrule
\multirow{2}{*}{\makecell[l]{Moderate wet\\(AC 0.5\,mm, $\mu{=}0.86$)}}
  & v7(weather aware model) & 77.9\% & 5.9\% & \textbf{27.8} \\
  & v8(weather blind model) & 78.0\% & 6.3\% & \textbf{58.7} \\
\bottomrule
\end{tabular}
\end{table}

SR is indistinguishable between policies ($\Delta\text{SR}\leq0.8$\,pp) --- a negative result for friction-aware navigation.
However, on moderate-wet road v7's mean peak DRAC drops from 37.1 to 27.8 while v8's rises from 50.2 to 58.7, a 2.1$\times$ gap absent on dry asphalt, consistent with the friction-aware policy learning gentler braking under reduced traction.
Full results including the hard-rain degenerate condition are in Appendix~\ref{app:friction_ablation}.

\section{Conclusion}
\label{sec:conclusion}

We presented \textbf{SceneFactory}, a GPU-accelerated, physics-grounded, and weather-aware multi-agent autonomous-driving simulation platform built on NVIDIA Isaac Sim and Isaac Lab.
SceneFactory procedurally converts Waymo Open Motion Dataset scenarios into GPU-resident USD road environments and parallelises up to 256 independent worlds on a single RTX PRO 6000 GPU, each populated with up to 16 articulated PhysX vehicles whose joint states, observations, rewards, and policy inference are all computed via batched tensor operations over the Isaac Lab GPU tensor API.
This design reaches 19,250 CASPS at 256 worlds $\times$ 16 agents, a 127$\times$ gain over a non-vectorized PhysX baseline using the same GPU and physics solver, making large-scale PPO training with full rigid-body dynamics practical on a single workstation GPU.
A CEM-based system identification procedure calibrates the 20-parameter vehicle model against a validated PhysX Vehicle teacher across 139 scripted maneuvers, and a physics-based weather-to-friction module maps precipitation and road surface type to per-world tire friction coefficients.

On 64 held-out Waymo scenes, the trained policy achieves 80.1\% goal-reach (500/624 agents) at 314M environment steps.
A vehicle dynamics ablation establishes that this 20\,pp gap relative to kinematic-bicycle simulators is attributable to the harder control problem posed by full rigid-body dynamics: a kinematic-bicycle-trained policy achieves 100\% SR in its own environment but collapses to 47.3\% when transferred to PhysX, while the PhysX-trained policy transfers back to the bicycle environment at 99.5\% SR.
A friction-conditioning experiment (§\ref{sec:ablation_friction}) found that a friction-aware policy did not outperform a friction-blind baseline at this training scale, which we report as a negative result and attribute to insufficient wet-world exposure and a coarse evaluation metric.

\paragraph{Limitations.}
Navigation success rates remain below those of kinematic-dynamics simulators; longer training budgets, reward curriculum refinement, and improved scene diversity are the most direct levers.
All road geometry is z-flattened, preventing simulation of multi-level road structures.
Per-world friction is spatially uniform; local surface variation (wet patches, surface transitions) is not representable in the current architecture.
Physical accuracy of the friction module has not been validated against measured braking or lateral-force data; it is best described as configurable and physically motivated rather than validated.

\paragraph{Future work.}
Several extensions are immediate priorities.
First, scaling to the full Waymo Open Motion Dataset (over 100{,}000 scenarios) is an engineering task and would substantially increase scene diversity.
Second, integrating formal near-miss metrics (time-to-collision distributions, post-encroachment time, and deceleration rate to avoid crash) as both training signals and evaluation metrics would align SceneFactory more closely with safety-critical assessment standards.
Third, dynamic within-episode weather variation (rain onset, drying) would test policy robustness to non-stationary friction.
Finally, coupling SceneFactory with NVIDIA Cosmos~\cite{nvidia_cosmos_2025} for photorealistic video generation from simulated trajectories would open a pathway from physics-based policy training to sensor-realistic evaluation.


\bibliographystyle{unsrtnat}
\bibliography{references}

\newpage

\appendix

\section{SceneFactory Pipeline}
\label{app:pipeline}

\textbf{Offline preprocessing} converts Waymo TFRecords into USD road assets (JSON manifest $\to$ USD prims $\to$ GPU tensors), run once and parallelised across scenarios.
\textbf{Environment instantiation} loads $N$ scenes into $W$ Isaac Lab worlds, spawns vehicles, and sets per-world friction via the weather model (\S\ref{sec:weather_friction}).
\textbf{Training loop} steps at 30\,Hz: gather GPU observations, infer actions, advance PhysX at 120\,Hz, compute rewards; after $T$ steps run PPO on $W{\times}M{\times}T$ transitions and reset all worlds.
Full pseudocode is given in Algorithm~\ref{alg:pipeline}.

\begin{algorithm}[h]
\caption{SceneFactory Pipeline}
\label{alg:pipeline}
\small
\begin{algorithmic}[1]
\Statex \textbf{Offline preprocessing} \hfill\Comment{run once per dataset}
\For{each Waymo TFRecord scenario $s$}
  \State Extract road polylines (lane centers, edges) and agent metadata $\to$ JSON
  \State Re-center coordinates; filter agents outside 400\,m bounding box
  \State Build USD road prims; bake polyline metadata into \texttt{customData}
  \State Save per-scenario USD and updated JSON manifest
\EndFor
\Statex
\Statex \textbf{Environment instantiation} \hfill\Comment{run at training startup}
\State Load scene pool of $N$ validated scenarios
\State Replicate scenes to fill $W$ parallel worlds (each world $\leftarrow$ one scenario)
\State Clone Isaac Lab template $W$ times on GPU; spawn articulated vehicles
\State Compute per-world friction $\mu \leftarrow$ \Call{WeatherToFriction}{surface, $h$}
\Statex
\Statex \textbf{Training loop} \hfill\Comment{repeated each PPO iteration}
\For{each rollout step $t = 1 \ldots T$}
  \State Observe: ego state, $k$-NN road points, neighbor vehicles, weather context
  \State $\mathbf{a}_t \leftarrow \pi_\theta(\mathbf{o}_t)$ \Comment{shared policy, on-device inference}
  \State Step PhysX at 120\,Hz (4 sub-steps per 30\,Hz control tick)
  \State Compute rewards; log goal-reach, collision, off-road events
  \State Reset agents that have terminated; park with zero velocity
\EndFor
\State Perform PPO update on $W \times M \times T$ transitions (on-device)
\State Reset all worlds with fresh start--goal assignments
\end{algorithmic}
\end{algorithm}

\section{Weather-to-Friction Model Details}
\label{app:friction_model}

This appendix provides the full mathematical formulation of the tire--pavement friction model used in SceneFactory (summarized in \S\ref{sec:weather_friction}).
We implement the modified Average Lumped LuGre (ALL) model~\cite{zhao_tire-pavement_2024}, which extends the LuGre bristle-deflection framework to account for pavement texture and water film thickness.

\paragraph{Bristle dynamics.}
The model evolves the average bristle deflection $\bar{z}$ via the ODE
\begin{equation}
\label{eq:app_bristle_ode}
\frac{d\bar{z}}{dt} = v_r - \theta \, Y_R \left[\frac{\sigma_0 \lvert v_r \rvert}{g(v_r)}\,\bar{z} + K \lvert \omega_r \rvert\,\bar{z}\right],
\end{equation}
where $v_r$ is the relative (slip) speed between the tire contact patch and the road, $\omega_r$ is the wheel circumferential speed, $\sigma_0$ is the bristle stiffness, $K = 7/(6L)$ is the ALL approximation constant with $L$ the contact patch length, $\theta$ is a texture influence coefficient (per road surface type), and $Y_R \in [0,1]$ is the contact patch length ratio under hydrodynamic lift.

\paragraph{Stribeck function.}
The steady-state function $g(v_r) = \mu_c + (\mu_s - \mu_c)\exp\bigl(-\lvert v_r / v_s \rvert^\alpha\bigr)$ captures the transition from static friction $\mu_s$ to Coulomb friction $\mu_c$.
The Stribeck speed depends on the water film thickness $h_w$ (in meters) via
\begin{equation}
\label{eq:app_stribeck_speed}
v_s(h_w) = b_1 \exp(1000\,b_2\,h_w + b_3) + b_4,
\end{equation}
where $b_1 = 4.8916$, $b_2 = -7.91$, $b_3 = 3.01$, and $b_4 = 3.40$ are calibrated constants from~\cite{zhao_tire-pavement_2024}.
Because $b_2 < 0$, $v_s$ decreases with water film thickness: thicker water films cause the friction drop-off to onset at lower slip speeds.

\paragraph{Effective friction coefficient.}
Given the steady-state solution $\bar{z}^*$, the effective friction coefficient is
\begin{equation}
\label{eq:app_mu_all}
\mu = \max\!\bigl(\theta\,Y_R - Y_F,\; 0\bigr) \cdot \bigl(\theta\,Y_R\,\sigma_0\,\bar{z}^* + \sigma_1\,\dot{\bar{z}} + \sigma_2\,v_r\bigr),
\end{equation}
where $Y_F \in [0,1]$ is the hydrodynamic force ratio and $\sigma_1$, $\sigma_2$ are viscous damping coefficients.
Under the default calibration from~\cite{zhao_tire-pavement_2024}, both $\sigma_1 = 0$ and $\sigma_2 = 0$, so the effective formula reduces to $\mu = \max(\theta\,Y_R - Y_F,\,0)\cdot\theta\,Y_R\,\sigma_0\,\bar{z}^*$.
Both $Y_R$ and $Y_F$ depend on vehicle speed, water film thickness, tire geometry (contact patch dimensions, tread radius, wheel radius), and physical constants (water density and viscosity).
When the contact factor $\theta\,Y_R - Y_F$ drops to zero, the model predicts full hydroplaning.
Our implementation numerically integrates Equation~\ref{eq:app_bristle_ode} to steady state using exponential integration ($\Delta t = 10^{-3}$\,s, convergence threshold $\epsilon = 10^{-7}$), matching the paper's calibrated parameters (Table~4 and Table~5 in~\cite{zhao_tire-pavement_2024}).

\begin{figure}[h]
\centering
\begin{tikzpicture}
  \begin{axis}[
    width=1\columnwidth,
    height=4.8cm,
    xlabel={Water-film depth (mm)},
    ylabel={Effective friction $\mu_{\mathrm{static}}$},
    xmin=0, xmax=0.88,
    ymin=0.75, ymax=1.45,
    xtick={0,0.2,0.4,0.6,0.8},
    ytick={0.8,0.9,1.0,1.1,1.2,1.3,1.4},
    tick label style={font=\small},
    label style={font=\small},
    grid=major,
    grid style={dashed,gray!35},
    legend style={font=\small, at={(0.95,0.95)}, anchor=north east},
    mark size=1.5pt,
    clip=false,
  ]
    \addplot[draw=none, fill=red!8, forget plot]
      coordinates {(0.82,0.75)(0.88,0.75)(0.88,1.45)(0.82,1.45)} \closedcycle;
    \node[font=\tiny, rotate=90, text=red!60] at (axis cs:0.855,1.08) {aquaplane};

    \draw[dashed, gray!60, thin] (axis cs:0.0,0.75) -- (axis cs:0.0,1.45);
    \draw[dashed, gray!60, thin] (axis cs:0.5,0.75) -- (axis cs:0.5,1.45);
    \node[font=\tiny, gray!70, anchor=south] at (axis cs:0.0,1.45) {dry};
    \node[font=\tiny, gray!70, anchor=south] at (axis cs:0.5,1.45) {mod.\ wet};

    \addplot[color=blue!70, mark=*, thick] coordinates {
      (0.0,1.1048)(0.1,1.0657)(0.2,1.0161)(0.3,0.9593)
      (0.4,0.9039)(0.5,0.8595)(0.6,0.8302)(0.7,0.8134)(0.8,0.8044)};
    \addlegendentry{AC}

    \addplot[color=orange!80!black, mark=triangle*, thick] coordinates {
      (0.0,1.2043)(0.1,1.1618)(0.2,1.1078)(0.3,1.0461)
      (0.4,0.9857)(0.5,0.9374)(0.6,0.9056)(0.7,0.8874)(0.8,0.8776)};
    \addlegendentry{SMA}

    \addplot[color=teal!80!black, mark=square*, thick] coordinates {
      (0.0,1.3368)(0.1,1.2898)(0.2,1.2301)(0.3,1.1617)
      (0.4,1.0948)(0.5,1.0413)(0.6,1.0061)(0.7,0.9860)(0.8,0.9753)};
    \addlegendentry{OGFC}

  \end{axis}
\end{tikzpicture}
\caption{Effective static friction $\mu_{\mathrm{static}}$ vs.\ water-film depth for the three road surface presets in SceneFactory, computed from the modified ALL model~\cite{zhao_tire-pavement_2024}.
OGFC retains the highest grip at all film depths due to its coarser texture ($\theta{=}1.21$, $T_d{=}1.08$\,mm) vs.\ SMA ($\theta{=}1.09$) and AC ($\theta{=}1.00$).
Vertical dashed lines mark the dry (0\,mm) and moderate-wet (0.5\,mm) eval conditions used in Appendix~\ref{app:friction_ablation}.
The shaded band marks the aquaplaning threshold (${{\gtrsim}}0.85$\,mm) beyond which the model predicts full grip loss.}
\label{fig:friction_mu_vs_film}
\end{figure}
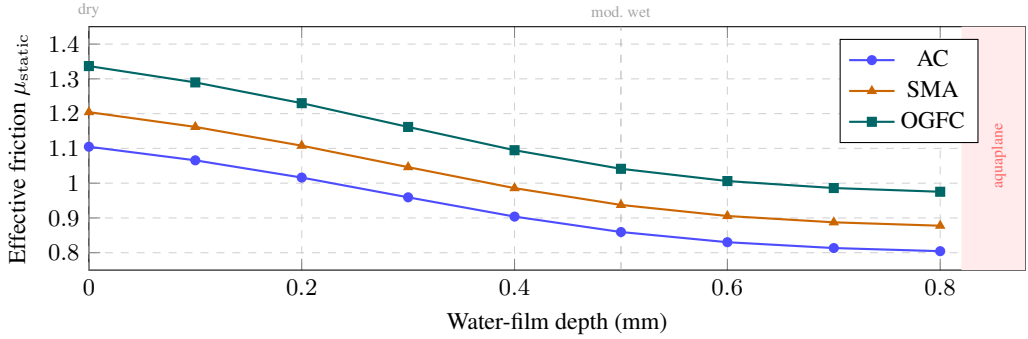

\paragraph{Road surface presets.}
The texture influence coefficient $\theta$ and the texture amplitude $T_d$ vary by pavement type.
We provide three presets calibrated to the values in~\cite{zhao_tire-pavement_2024}:
\emph{Asphalt Concrete} (AC, $\theta{=}1.00$, $T_d{=}0.65$\,mm),
\emph{Stone Mastic Asphalt} (SMA, $\theta{=}1.09$, $T_d{=}0.80$\,mm), and
\emph{Open-Graded Friction Course} (OGFC, $\theta{=}1.21$, $T_d{=}1.08$\,mm).
OGFC, designed for drainage, retains higher friction under wet conditions; AC, the most common pavement, serves as the baseline.
Figure~\ref{fig:friction_mu_vs_film} plots the resulting $\mu_{\mathrm{static}}$ curves against water-film depth for all three presets.

\begin{table}[h]
\centering\small
\caption{Static friction coefficient $\mu$ from the ALL model~\cite{zhao_tire-pavement_2024} at 13.89\,m/s for three road surface types and five water film thicknesses $h$. Values at $h\geq1.0$\,mm indicate full hydrodynamic lift; the simulator applies a floor of $10^{-3}$.}
\label{tab:friction_mu}
\begin{tabular}{lccccc}
\toprule
Surface & $h=0$ (dry) & $h=0.3$\,mm & $h=0.5$\,mm & $h=1.0$\,mm & $h=2.0$\,mm \\
\midrule
AC   & 1.105 & 0.959 & 0.859 & $\leq 10^{-3}$ & $\leq 10^{-3}$ \\
SMA  & 1.204 & 1.046 & 0.937 & $\leq 10^{-3}$ & $\leq 10^{-3}$ \\
OGFC & 1.337 & 1.162 & 1.041 & $\leq 10^{-3}$ & $\leq 10^{-3}$ \\
\bottomrule
\end{tabular}
\end{table}

\paragraph{PhysX integration pipeline.}
Each world in the training stage is assigned a friction configuration specifying the road surface type and the water film thickness $h_w$ (in mm).
At environment construction time, the pipeline feeds these values into the modified ALL model (Equation~\ref{eq:app_mu_all}) at a reference speed of 13.89\,m/s ($\approx$50\,km/h) and two slip ratios (0.15 for static, 0.80 for dynamic) to produce per-world static and dynamic friction coefficients $\mu_s$ and $\mu_d$.
The effective coefficient $\mu_{\mathrm{eff}}$ (static by default) is used downstream as follows.

The ground-plane material uses the sysid-tuned friction scale for the dry-asphalt surface type (Section~\ref{sec:vehicle_sysid}), with static friction $\mu^{\mathrm{ground}}_s = \min(1.0,\; f_{\mathrm{surface}} \cdot \sqrt{f_{\mathrm{lon}} \cdot f_{\mathrm{lat}}})$ and dynamic friction $\mu^{\mathrm{ground}}_d = 0.95\,\mu^{\mathrm{ground}}_s$.
PhysX resolves contact friction using the minimum of the ground and tire material values (combine mode \texttt{min}).
Because all cloned environments share a single global ground plane, this material is uniform across worlds.
Per-world friction variation is currently communicated to the policy only through the observation (Equation~\ref{eq:weather_obs}).

\begin{figure}
    \centering
    \includegraphics[width=1\linewidth]{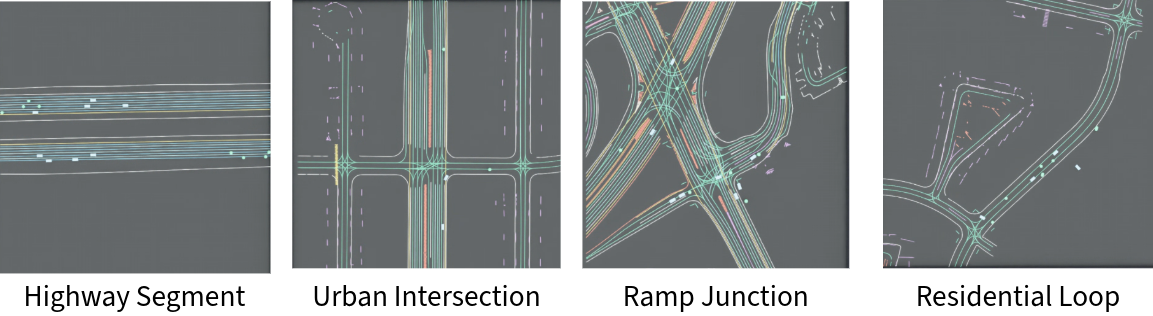}
    \caption{Randomly loaded Scenefactory scenes with diverse road topologies.}
    \label{fig:comp_sc}
\end{figure}
\section{Scene Construction Details}
\label{app:scene_construction}

\paragraph{Per-scenario world building steps.}
Given a scene JSON, the world builder constructs USD road geometry through five steps:
\begin{enumerate}[leftmargin=*, itemsep=1pt]
  \item \textbf{Re-centering.}
  A scene center is computed as the mean of all road polyline points.
  All coordinates (road and agent) are shifted so that each scenario is locally centered at the origin.

  \item \textbf{Z-flattening.}
  Waymo road elevation data is collapsed to a constant ground height ($z=0$).
  Scenes with significant 3D structure (bridges, ramps) are filtered during curation (Appendix~\ref{app:sharp_edges}).

  \item \textbf{Segment generation.}
  Each consecutive pair of polyline points defines an oriented box segment placed at the midpoint and rotated to align with the point-to-point direction (width 0.10\,m, height 0.10\,m).
  Pairs separated by more than 3.0\,m are treated as discontinuities and skipped.
  Both endpoints must fall within $\pm B/2$ of the scene center ($B{=}200$\,m).

  \item \textbf{USD prim creation.}
  Segments are grouped by polyline type and rendered as a \texttt{PointInstancer} with a single unit-cube prototype; per-segment position, orientation, and scale arrays encode the geometry compactly. The different reconstructed Waymo Motion Open Dataset scene is illustrated in Figure \ref{fig:comp_sc}.

  \item \textbf{Metadata baking.}
  Five arrays are written to the world root prim's USD \texttt{customData}: segment midpoint positions, direction vectors, polyline type codes, half-lengths, and half-widths.
  This flat, GPU-friendly format enables constant-time runtime access without re-parsing the JSON or traversing the USD prim hierarchy.
\end{enumerate}

\begin{figure}
    \centering
    \includegraphics[width=1\linewidth]{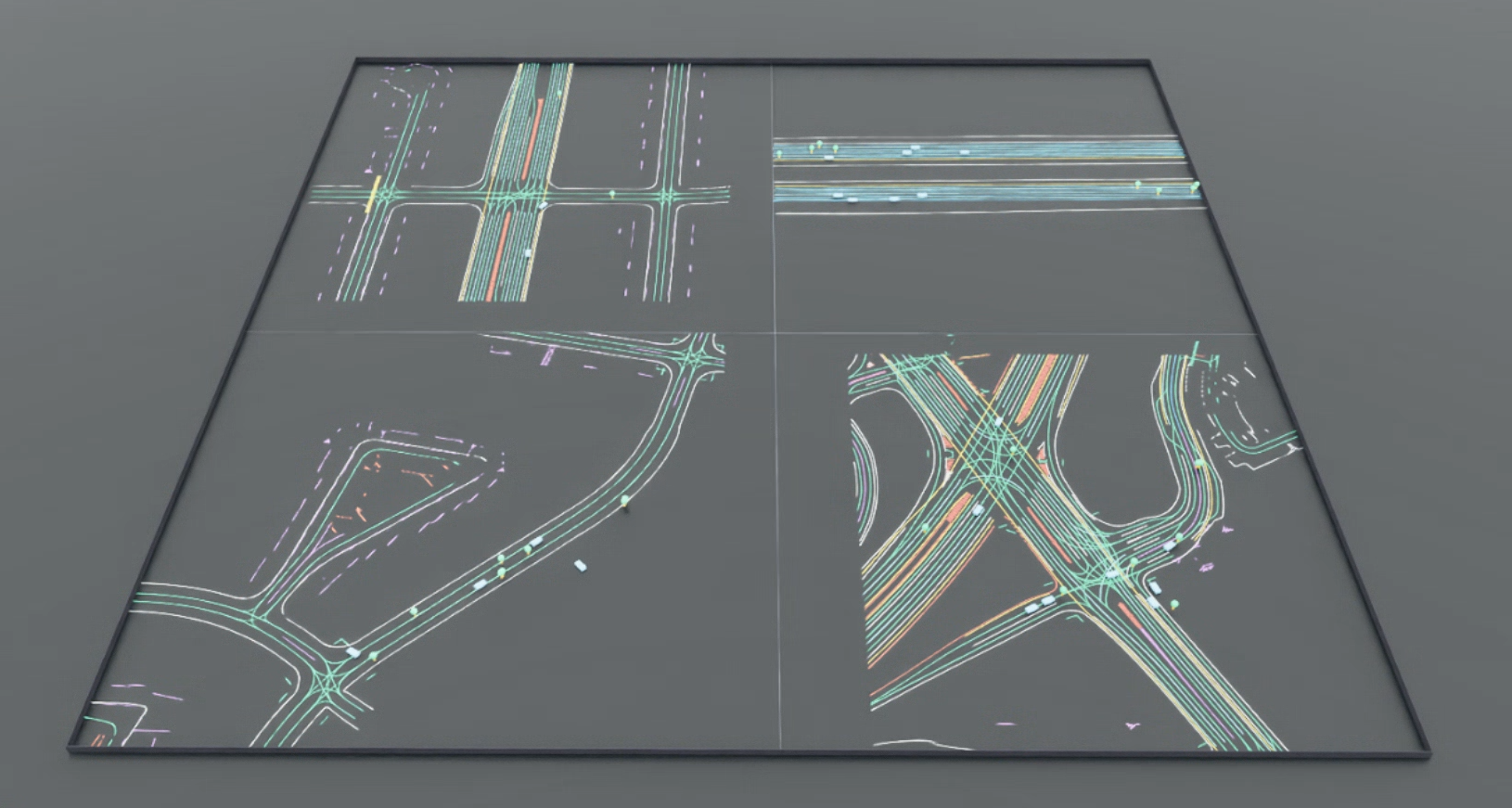}
    \caption{Scenefactory loaded with four randomly selected Waymo Open Motion Dataset traffic scene. }
    \label{fig:multi_world}
\end{figure}

\paragraph{Multi-world grid assembly.}
SceneFactory arranges $N$ worlds in a 2D grid, each occupying a $200 \times 200$\,m cell with 200\,m padding (400\,m pitch between centers).
In \emph{random-fill} mode, the scene pool is shuffled (seed~42) and tiled to fill $N$ slots; in \emph{fixed} mode all environments share the same scenario. A screenshot of SceneFactory when loaded with 4 parallel worlds is shown in Figure \ref{fig:multi_world}
\paragraph{Post-clone road injection.}
Isaac Lab's cloning mechanism replicates a single template environment $N$ times.
SceneFactory builds road geometry \emph{after} the clone step, so the template contains only vehicle articulations and the ground plane, and each environment can host a different Waymo-derived layout without cloning complex instancer prims (which is unreliable under Fabric).

\paragraph{GPU metadata tensors.}
At startup, the baked \texttt{customData} arrays are loaded into padded GPU tensors of shape $(N, P_{\max}, d)$, where $P_{\max}$ is the maximum segment count across all environments.
A boolean validity mask distinguishes real segments from padding, supporting vectorized lane-proximity queries and oriented-box overlap tests during training with no USD access in the inner loop.

\section{Observation Space Details}
\label{app:observations}

Each agent receives a 1{,}929-dimensional observation vector assembled from four groups.
All spatial quantities are expressed in the agent's body frame via a 2D rotation by $-\psi$ (ego heading), so the policy sees an ego-centric view of the world.

\paragraph{Ego state (7-dim).}
The ego state encodes the agent's goal and velocity in body-frame coordinates:
\begin{enumerate}[leftmargin=*, itemsep=1pt]
  \item Goal position $(g_x^b, g_y^b)$, each divided by the scene bounding-box half-size ($B/2 = 100$\,m).
  \item Heading error to the goal, encoded as $(\sin\Delta\psi,\,\cos\Delta\psi)$.
  \item Euclidean goal distance $\|g^b_{xy}\|$, divided by $B/2$.
  \item Body-frame velocity $(v_x^b, v_y^b)$, each divided by 10\,m/s.
\end{enumerate}
When the weather-to-friction module is active, the 4-dim weather context (Equation~\ref{eq:weather_obs}) is appended to the ego group, yielding an 11-dim input to the ego encoder.

\paragraph{Road geometry context ($K_r \times 5 = 1{,}750$-dim).}
Each agent selects $K_r = 350$ road-segment midpoints from the pre-loaded GPU metadata tensors (\S\ref{sec:data_processing}).
The selection operates in two steps: first, all road points within a radius $r = 10$\,m of the ego position are identified via a squared-distance threshold; second, these candidate points are sorted by their original polyline index to preserve road topology, and the first $K_r$ are taken.
If fewer than $K_r$ points fall within the radius, the remaining slots are zero-padded.
Each selected point contributes five features:
\begin{enumerate}[leftmargin=*, itemsep=1pt]
  \item Relative position $(\Delta x^b, \Delta y^b)$ in ego body frame, each divided by $r$.
  \item Polyline type code, divided by a normalizer ($= 20$).
  \item Road direction unit vector $(d_x^b, d_y^b)$ in ego body frame.
\end{enumerate}

\paragraph{Neighbor vehicles ($K_v \times 7 = 168$-dim).}
The $K_v = 24$ nearest alive agents (by Euclidean distance in world XY) are selected per ego agent.
Self-distance and dead-agent distances are set to infinity before the argsort.
Each neighbor contributes seven features:
\begin{enumerate}[leftmargin=*, itemsep=1pt]
  \item Relative position $(\Delta x^b, \Delta y^b)$ in ego body frame, each divided by $B/2$.
  \item Chassis length and width, each divided by $B/2$.
  \item Relative heading $\Delta\psi$, divided by $\pi$.
  \item Speed $\|v_{xy}\|$, divided by 10\,m/s.
  \item Pairwise time-to-collision (TTC), computed via a swept-circle test.
    Each vehicle is approximated by three equal-radius circles placed along the longitudinal axis at offsets $\{-d,\, 0,\, +d\}$ from the chassis centre, where $r = \max(0.45,\; 0.55\, W)$ and $d = \min\!\bigl(\tfrac{\mathit{WB}}{2},\; \max(0,\; \tfrac{L}{2} - 0.8\,r)\bigr)$.
    For the default vehicle ($L{=}4.0$\,m, $W{=}2.0$\,m, $\mathit{WB}{=}2.6$\,m) this gives $r \approx 1.10$\,m and $d \approx 1.12$\,m.
    The test solves a quadratic closest-approach equation for every pair among the $3 \times 3 = 9$ circle combinations and takes the minimum.
    The resulting TTC is clamped to $[0,\, \tau_{\max}]$ and divided by $\tau_{\max} = 10$\,s.
    Neighbors behind the ego or with no collision trajectory receive $\mathrm{TTC}/\tau_{\max} = 1$.
\end{enumerate}
Slots beyond the number of alive neighbors are zero-padded.

\section{Network Architecture Details}
\label{app:network}

The observation groups are processed by a late-fusion actor-critic network with per-modality encoders and max-pool set aggregation, following the design of GPUDrive~\cite{kazemkhani_gpudrive_2025}.
We make two modifications: (i)~we use \emph{masked} max-pool (zero-padded slots are filled with $-\infty$ before the $\max$) rather than a plain pool, and (ii)~we replace their Tanh activations with ELU.
The flat observation is split into its three groups and processed by separate encoder MLPs, one set for the actor and one for the critic (no weight sharing).

\begin{itemize}[leftmargin=*, itemsep=1pt]
  \item \textbf{Ego encoder}: $11 \to 64 \to 64$ (two hidden layers, ELU activations).
  \item \textbf{Road encoder}: $5 \to 96 \to 96$, applied independently to each of the $K_r$ points; the $K_r$ output vectors are aggregated via \emph{masked max-pool}, where zero-padded points are filled with $-\infty$ before the $\max$ operation.
  \item \textbf{Vehicle encoder}: $7 \to 96 \to 96$, applied independently to each of the $K_v$ neighbors; aggregated via masked max-pool in the same way.
\end{itemize}

The three embeddings (64 + 96 + 96 = 256) are concatenated and passed through a shared trunk MLP ($256 \to 128 \to 64$, ELU).
The actor head projects to a 3-dim mean vector $\boldsymbol{\mu}$ (one per action), paired with a state-independent learnable log-standard-deviation $\log\boldsymbol{\sigma}$ to form a diagonal Gaussian policy.
The critic head projects to a scalar state-value estimate.
No empirical observation normalization (running mean/std) is used; all features are manually scaled at assembly time as described in Appendix~\ref{app:observations}.

\section{Action Space Details}
\label{app:actions}

Each agent outputs a three-dimensional continuous action $\mathbf{a} = (a_{\mathrm{thr}},\, a_{\mathrm{str}},\, a_{\mathrm{brk}})$.
The policy produces values in $[-1,1]^3$; throttle and brake are rectified to $[0,1]$ before decoding, so a zero-mean Gaussian policy naturally idles.
All three channels are decoded into PhysX \emph{effort targets}; the steering PD loop described in Eq.\ref{eq:steer_pd} is computed in software at every physics substep.

\paragraph{Control frequency.}
Physics runs at $1/\Delta t_{\mathrm{phys}} = 120$\,Hz.
Actions are held constant over a decimation window of 4 substeps, yielding a control frequency of 30\,Hz.
Within each substep the steering PD reads the current joint state, so the effective PD loop rate is 120\,Hz.

\paragraph{Brake sign memory.}
Brake torque must oppose wheel rotation, so its sign is $-\mathrm{sign}(\dot{q}_{\mathrm{wheel}})$.
When $|\dot{q}_{\mathrm{wheel}}| < 10^{-4}$\,rad/s the sign becomes unreliable; we therefore latch the last observed sign and continue applying it until rotation resumes.
This prevents torque chatter near zero velocity.

\paragraph{Actuation parameters.}
Table~\ref{tab:app_action_params} lists the sysid-tuned values used in all experiments.
Drive torque is applied to the two front wheels only (FWD); brake torque is applied to all four wheels with separate front/rear magnitudes.
The sysid procedure additionally fits per-surface friction scales (longitudinal and lateral) that modulate the effective contact friction for each road surface type; these are relevant only for multi-surface scenarios (\S\ref{sec:weather_friction}).

\begin{table}[h]
  \centering\small
  \caption{Action-decoding parameters (sysid-tuned values).}
  \label{tab:app_action_params}
  \begin{tabular}{@{}lll@{}}
    \toprule
    Parameter & Symbol & Value \\
    \midrule
    Max steer angle        & $\theta_{\max}$               & $0.450$\,rad ($\approx 25.8^{\circ}$) \\
    Steering PD $K_p$      & $K_p$                          & $1839.5$\,N$\cdot$m/rad \\
    Steering PD $K_d$      & $K_d$                          & $110.5$\,N$\cdot$m$\cdot$s/rad \\
    Steering effort clamp  & $\tau_{\mathrm{steer,max}}$   & $1200$\,N$\cdot$m \\
    Drive torque           & $\tau_{\mathrm{drive,max}}$   & $600.7$\,N$\cdot$m \\
    Brake torque (front)   & $\tau_{\mathrm{brk,F}}$      & $1090.5$\,N$\cdot$m \\
    Brake torque (rear)    & $\tau_{\mathrm{brk,R}}$      & $980.7$\,N$\cdot$m \\
    Wheel mass             & $m_{\mathrm{wheel}}$          & $37.5$\,kg \\
    Wheel inertia scale    & $s_{\mathrm{inertia}}$        & $1.094$ \\
    Suspension stiffness   & $k_{\mathrm{susp}}$           & $1080.8$\,N/m \\
    Suspension damping      & $c_{\mathrm{susp}}$           & $2764.3$\,N$\cdot$s/m \\
    Yaw stability damping  & $\lambda_{\mathrm{yaw}}$      & $10.6$\,N$\cdot$m$\cdot$s/rad \\
    \bottomrule
  \end{tabular}
\end{table}

\section{Reward Design Details}
\label{app:rewards}

The per-step reward is the sum of ten terms: four sparse event rewards that also trigger agent termination, and six dense shaping rewards applied at every active step.
Tuning these weights required extensive iteration; the values reported here correspond to the configuration used in all experiments.

\paragraph{Sparse event rewards.}
Four mutually exclusive events terminate an agent and deliver a one-shot reward:
\begin{enumerate}[leftmargin=*, itemsep=1pt]
  \item \textbf{Goal reach} ($+45.0$).
  Awarded when the ego position is within $d_{\mathrm{goal}} = 3.0$\,m of the goal in XY.

  \item \textbf{Collision} ($-6.0$).
  Triggered when the maximum contact-sensor force exceeds 25\,N.
  A warm-up window of 24 steps after spawn suppresses false positives from initial settling.

  \item \textbf{Crash} ($-10.0$).
  Triggered when the vehicle falls below the ground plane ($z_{\mathrm{rel}} < 0$), drifts more than 100\,m from its start position, or tilts beyond a gravity-vector threshold ($g^b_z > -0.15$, indicating a rollover or severe pitch).

  \item \textbf{Lane-forbidden} ($-20.0$).
  Triggered when any of the vehicle's three bounding circles overlaps a road-edge segment (Waymo types 15--16), detected via an oriented-bounding-box overlap test.
\end{enumerate}
After termination, the agent is teleported off-stage and all subsequent dense rewards are masked to zero.

\paragraph{Dense shaping rewards.}
Six terms provide per-step signal for active agents:
\begin{enumerate}[leftmargin=*, itemsep=1pt]
  \item \textbf{Route progress} (weight $w = 2.0$).
  The displacement between consecutive steps is projected onto the tangent of the nearest driving-lane segment (Waymo types 1--2), oriented toward the goal:
  $r_{\mathrm{prog}} = \mathrm{clamp}\bigl((\mathbf{p}_t - \mathbf{p}_{t-1}) \cdot \hat{\mathbf{t}}_{\mathrm{lane}},\; -2,\; +2\bigr) \cdot w$.
  This rewards forward progress along the route and penalizes backwards motion.

  \item \textbf{Geometric lane-keeping} (weight $w = 0.08$).
  A continuous quality score combines lateral offset and heading alignment with the nearest driving lane:
  \[
    q = \exp\!\bigl(-(\ell / \sigma)^2\bigr) \cdot \bigl[(1 - w_h) + w_h \cdot \max(0,\, \cos(\psi - \psi_{\mathrm{lane}}))\bigr],
  \]
  where $\ell$ is the signed lateral offset, $\sigma = 1.75$\,m is the tolerance, and $w_h = 0.8$ is the heading weight.
  The reward is $r_{\mathrm{lane}} = q \cdot w$.

  \item \textbf{Off-road penalty} (weight $w = -0.5$).
  Applied when the vehicle's lateral offset exceeds 3.25\,m or its distance to the nearest lane exceeds 6.0\,m.

  \item \textbf{Idle penalty} (weight $w = -0.05$).
  Applied when the vehicle speed drops below 1.0\,m/s, discouraging the policy from learning to stop and wait.

  \item \textbf{Inter-vehicle TTC penalty.}
  The minimum pairwise TTC across all neighbors is converted to a penalty:
  \[
    p(\tau) = \min\!\bigl(\alpha_v / \max(\tau,\, \tau_{\mathrm{floor}}),\; p_{\mathrm{max},v}\bigr),
  \]
  with $\alpha_v = 0.10$, $p_{\mathrm{max},v} = 0.35$, and $\tau_{\mathrm{floor}} = 0.5$\,s.
  The reward is $r_{\mathrm{ttc}} = -p(\tau)$.

  \item \textbf{Road-edge TTC penalty.}
  The same $\alpha/\tau$ penalty form is applied to the closest road-edge segment within 40\,m ahead:
  $\alpha_e = 0.07$, $p_{\mathrm{max},e} = 0.40$, $\tau_{\mathrm{floor}} = 0.5$\,s.
  This encourages the policy to slow down when approaching road boundaries.
\end{enumerate}

\begin{table}[t]
  \centering\small
  \caption{Reward components used in all experiments.  Sparse rewards are delivered once and trigger agent termination; dense rewards are applied at every active step.}
  \label{tab:app_reward_terms}
  \begin{tabular}{@{}llrll@{}}
    \toprule
    Term & Type & Weight & Sign & Key threshold \\
    \midrule
    Goal reach         & Sparse & $45.0$  & $+$ & dist $\le 3.0$\,m \\
    Collision          & Sparse & $6.0$   & $-$ & force $\ge 25$\,N, 24-step warmup \\
    Crash              & Sparse & $10.0$  & $-$ & fall / drift $> 100$\,m / tilt \\
    Lane-forbidden     & Sparse & $20.0$  & $-$ & road-edge OBB overlap \\
    \midrule
    Route progress     & Dense  & $2.0$   & $\pm$ & clamp $[-2,+2]$\,m/step \\
    Lane-keeping       & Dense  & $0.08$  & $+$ & $\sigma{=}1.75$\,m, $w_h{=}0.8$ \\
    Off-road           & Dense  & $0.5$   & $-$ & lat $> 3.25$\,m or dist $> 6.0$\,m \\
    Idle               & Dense  & $0.05$  & $-$ & speed $< 1.0$\,m/s \\
    Inter-vehicle TTC  & Dense  & $\alpha{=}0.10$ & $-$ & max $= 0.35$, floor $= 0.5$\,s \\
    Road-edge TTC      & Dense  & $\alpha{=}0.07$ & $-$ & max $= 0.40$, floor $= 0.5$\,s \\
    \bottomrule
  \end{tabular}
\end{table}

\paragraph{Termination and time-out.}
An agent terminates upon any sparse event (goal, collision, crash, or lane-forbidden).
If no event occurs, the episode ends at $T_{\mathrm{max}} = 50$\,s (1{,}500 steps at 30\,Hz).
Terminated agents are teleported to an off-stage location ($x = 1{,}000$\,m) and excluded from subsequent reward and observation computation.

\section{System Identification Details}
\label{app:sysid}
\begin{figure}
    \centering
    \includegraphics[width=1\linewidth]{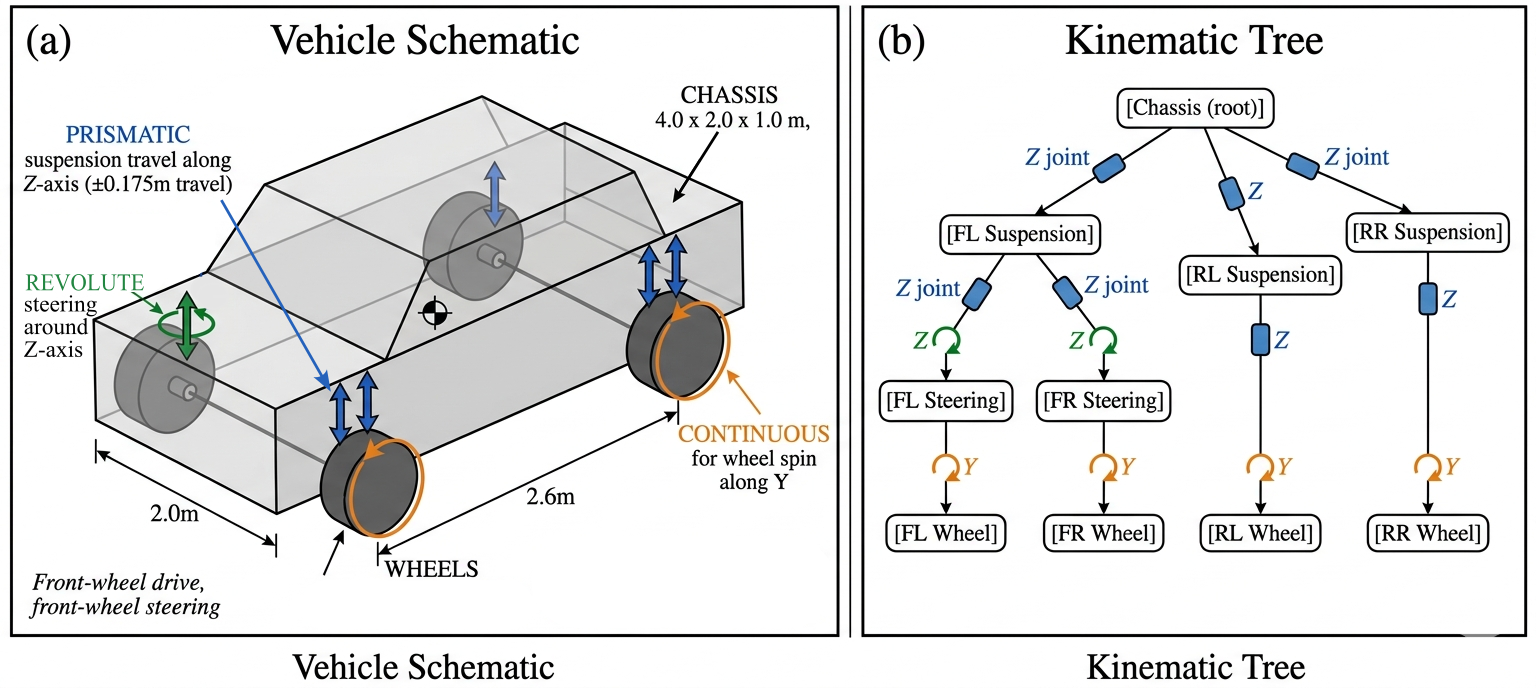}
    \caption{Vehicle model configuration used in Scenefactory. (The non-rectangular upper body is included only for visual illustration; the actual vehicle body is modeled as a rectangular cuboid.)}
    \label{fig:vehicle}
\end{figure}
The vehicle model (shown in Figure \ref{fig:vehicle}) has 20 tunable dynamics parameters: drive, brake (front/rear), and steering torques; steering PD gains and limits; suspension stiffness and damping; wheel mass and inertia scale; chassis center-of-mass offset; lateral and yaw damping coefficients; and per-surface friction scales for three surface types (dry, wet, gravel).
We fit these parameters using the cross-entropy method (CEM).

\paragraph{Reference data.}
A PhysX ``teacher'' vehicle (the Isaac Sim built-in PhysX Vehicle model, which includes a validated tire model, drivetrain, and suspension) executes 139 scripted maneuvers on a multi-surface patch track.
The maneuvers span five tiers: 17 longitudinal (throttle sweep 10--100\%, brake sweep, linear ramps), 64 lateral (step-steer and constant-radius at four throttle levels $\times$ four--five steer amplitudes $\times$ left/right), 18 combined (trail-brake at three throttle$\times$steer$\times$direction), 39 frequency-response (sinusoidal steering at multiple throttle/amplitude/frequency settings, plus chirp sweeps), and 1 surface transition across dry asphalt ($\mu_s{=}1.0$), wet asphalt ($\mu_s{=}0.75$), and gravel ($\mu_s{=}0.60$).
The comprehensive suite covers the full operating envelope from 10\% to 100\% throttle, ensuring that the fitted parameters generalize across the vehicle's speed and lateral-force range.
Each rollout records per-frame chassis position, yaw, speed, yaw rate, wheel angular velocities, and steer angles at 60\,Hz.

\paragraph{Loss function.}
Let $x^{\mathrm{ref}}_{k,t}$ denote the teacher's value for channel $k$ at frame $t$, and $x^{\mathrm{stu}}_{k,t}$ the corresponding student value under a candidate parameter vector.
Each rollout has $T$ frames recorded at 60\,Hz.
The per-channel mean squared error is $\mathrm{MSE}_k = \tfrac{1}{T}\sum_{t=1}^{T}(x^{\mathrm{stu}}_{k,t} - x^{\mathrm{ref}}_{k,t})^2$.
The full objective adds two terminal-frame penalties that emphasize end-of-maneuver accuracy:
\begin{equation}
\label{eq:app_sysid_loss}
\mathcal{L} = \sum_{k \in \mathcal{K}} w_k \cdot \mathrm{MSE}_k
  + w_{\mathrm{pos,final}} \cdot \|\mathbf{p}^{\mathrm{stu}}_T - \mathbf{p}^{\mathrm{ref}}_T\|^2
  + w_{\mathrm{spd,final}} \cdot (s^{\mathrm{stu}}_T - s^{\mathrm{ref}}_T)^2,
\end{equation}
where $\mathcal{K} = \{\text{position}_{xy}, \text{yaw}, \text{speed}, \text{yaw rate}, \text{wheel speed}, \text{steer angle}\}$ with weights $w_k = (1.0, 0.4, 0.35, 0.25, 0.05, 0.05)$ respectively.
The vectors $\mathbf{p}^{\mathrm{stu}}_T, \mathbf{p}^{\mathrm{ref}}_T \in \mathbb{R}^2$ are the student and teacher XY positions at the final frame, and $s^{\mathrm{stu}}_T, s^{\mathrm{ref}}_T$ are the corresponding speeds.
The terminal weights are $w_{\mathrm{pos,final}} = 1.5$ and $w_{\mathrm{spd,final}} = 0.75$.

\paragraph{Multi-stage CEM search.}
The CEM search is divided into five stages to manage the 20-dimensional space:
\begin{enumerate}[leftmargin=*, itemsep=1pt]
  \item \emph{Longitudinal}: tunes drive/brake torques and wheel parameters on the straight-line rollout.
  \item \emph{Steering}: tunes steering gains, suspension, and damping on the turning rollouts.
  \item \emph{Surface}: tunes per-surface friction scales on the transition rollout.
  \item \emph{Refinement}: tunes all 20 parameters jointly on all rollouts with a narrowed search window (18\% of the full range).
  \item \emph{Brake preservation}: re-tunes longitudinal parameters with a tight window (10\%) to prevent regression.
\end{enumerate}
Each stage inherits the best configuration from the previous stage.
The CEM uses a population of 24 candidates, an elite fraction of 25\%, an initial standard deviation of 25\% of each parameter's range, and a minimum standard deviation of 5\% to prevent premature convergence.
A total of 320 trials are allocated across the five stages with weights (30\%, 20\%, 15\%, 20\%, 15\%).
Table~\ref{tab:sysid_stages} reports the per-stage results.

\begin{table}[h]
\centering\small
\caption{Per-stage sysid results from the comprehensive CEM run (320 trials, 139 teacher maneuvers).
``Best loss'' is the lowest $\mathcal{L}$ (Eq.\,\ref{eq:app_sysid_loss}) achieved in that stage; stages are cumulative so each inherits the best prior configuration.}
\label{tab:sysid_stages}
\begin{tabular}{@{}lrrrl@{}}
\toprule
Stage & Maneuvers & Trials & Best loss & Key parameters tuned \\
\midrule
Longitudinal     & 17  & 96 & 5.29  & drive torque, brake, wheel mass/inertia \\
Steering         & 121 & 64 & 170.6 & steering PD, suspension, yaw damping \\
Surface          & 1   & 48 & 55.4  & per-surface friction scales \\
Refinement       & 139 & 64 & 150.2 & all 20 parameters (narrow window) \\
Brake preserv.   & 17  & 48 & 3.69  & longitudinal re-anchor \\
\bottomrule
\end{tabular}
\end{table}

\paragraph{Vehicle-intrinsic vs.\ environment parameters.}
The 20 sysid parameters are \emph{vehicle-intrinsic}: they characterize how the articulation converts actuator commands into motion (drive torque, steering response, suspension compliance, etc.).
They do not change when the road surface changes.
Environment-level friction variation is handled orthogonally by the PhysX contact solver: each environment's wheel colliders are assigned a friction coefficient $\mu$ via their physics material, and the contact solver computes friction forces at the wheel--ground contact points accordingly (using combine mode \texttt{min}).
This separation is deliberate: sysid produces a realistic vehicle model once, and the physics engine then automatically adapts that vehicle's behavior to any friction environment.

\section{Simulator Limitations and Sharp Edges}
\label{app:sharp_edges}

Transparency about simulator assumptions is essential for interpreting the experimental results.
Below we catalog the most consequential simplifications, known failure modes, and practical workarounds in SceneFactory, organized by subsystem.

\paragraph{Z-flattening.}
The Waymo Open Motion Dataset provides full 3D polyline coordinates, including road elevation changes across bridges, ramps, overpasses, and multi-level interchanges.
SceneFactory collapses all road geometry to a constant ground height ($z{=}0$), because vehicles drive on a single flat PhysX ground plane per world.
This design decision has two consequences.
First, scenes with elevated structures (bridges, highway overpasses) produce visually overlapping road segments at ground level, which can confuse the $k$-nearest-road-point observation by presenting the agent with road points from an overhead or underpass road that it cannot physically reach.
Second, ramp grades and banked curves are lost, removing a class of driving challenges (hill starts, crest reactions) that affect real-world vehicle dynamics.
We mitigate the first issue through scene curation and by filtering road points whose original Waymo $z$ coordinate deviates significantly from the ground plane, but the fundamental limitation remains: SceneFactory is a 2.5D simulator operating on 3D data.

\paragraph{Waymo road-edge labeling.}
SceneFactory treats Waymo polyline types 15 and 16 as forbidden road edges for both the lane-forbidden termination event and the road-edge TTC penalty.
In practice, these labels are not uniformly reliable across the dataset.
As noted by Kazemkhani et al.~\cite{kazemkhani_gpudrive_2025}, the Waymo dataset labels parking-lot entrances as road edges.
Some tracked vehicles in these scenarios have their origin on the main road and their destination inside the parking area (or vice versa), so the recorded trajectory necessarily crosses a road-edge polyline.
When such a scenario is used for training, the agent faces an impossible conflict: reaching the goal requires crossing a boundary that triggers the lane-forbidden penalty ($-20.0$) and immediate termination.
These cases inject contradictory supervision into the training signal, degrading both goal-reach rate and lane-safety metrics.
Other scenarios lack road-edge annotations entirely where a physical curb or barrier clearly exists, creating blind spots in the lane-safety reward.
Scene curation removes the worst cases, but residual label noise remains and likely contributes to the modest lane-forbidden violation rates observed in evaluation.

\paragraph{Moving-only agent assumption.}
The spawn filter retains only Waymo agents whose start and goal positions are separated by more than 3.0\,m (the goal-reached threshold).
Agents below this threshold are treated as ``already at goal'' and excluded entirely.
Because the Waymo dataset records start and end positions from the logged trajectory, vehicles that remain parked, idle, or nearly stationary throughout the 9.1\,s recording window have near-zero start--goal displacement and are filtered out.
As a result, \emph{parked vehicles do not exist in the simulation}: a scene in which a moving vehicle must navigate past a row of curbside-parked cars will contain only the moving vehicle.
This simplification removes an entire class of realistic interactions (door-zone avoidance, double-parked obstacle negotiation, parallel parking maneuvers) and inflates the effective lane width available to the agent.
A future extension could retain filtered agents as static rigid-body obstacles rather than discarding them.

\paragraph{PointInstancer instability under Fabric.}
Road segments are rendered as USD \texttt{PointInstancer} prims for compactness (one instancer per road type per world, with per-instance position, orientation, and scale arrays).
We discovered that PointInstancer prims interact poorly with Isaac Lab's Fabric state management layer: visibility updates and prototype-table initialization are unreliable, producing intermittent rendering artifacts (invisible or misplaced road segments) that do not affect physics but corrupt visual debugging and video captures.
An explicit-prim fallback mode is provided (one \texttt{UsdGeom.Cube} per road segment) for diagnostic and stage-export workflows.
Production training disables Fabric entirely (\texttt{use\_fabric: false}) to avoid related issues with road metadata access, at a modest throughput cost.

\paragraph{Collision-sensor warmup.}
When vehicles are spawned or teleported to their start poses, PhysX requires several substeps to settle the suspension and establish stable wheel--ground contact.
During this transient, contact sensors report large spurious forces as the chassis drops onto the ground plane.
Without mitigation, these forces trigger false collision terminations at the start of every episode.
We suppress collision detection for the first 24 environment steps (${\approx}0.8$\,s) after each spawn, which is sufficient for the suspension to settle.
This warmup window introduces a brief period at episode start during which actual inter-vehicle collisions are unpenalized, though in practice vehicles are spawned with sufficient separation that early collisions are rare.

\paragraph{Teleport-only reset.}
Isaac Lab's default environment reset tears down and reconstructs PhysX actors, which is prohibitively expensive for multi-agent scenes with per-environment road geometry.
SceneFactory uses a \emph{teleport-only} reset strategy: after a single full initialization reset, subsequent episode resets reposition vehicles by writing directly to root-state tensors (position, orientation, linear and angular velocity) without rebuilding the physics scene.
This is ${\sim}10{\times}$ faster but has a subtle limitation: accumulated PhysX internal state (joint impulse caches, contact-pair broadphase data) is not fully cleared.
In rare cases this causes a vehicle to retain a residual velocity or contact state from the previous episode for one or two substeps after teleport.
The collision warmup window (above) masks most such artifacts.

\paragraph{Uniform friction surface.}
Per-world friction variation is implemented by writing to each vehicle's wheel collider physics material at environment construction time.
Because the global ground plane has friction $\mu{=}1.0$ and PhysX uses combine mode \texttt{min}, the effective contact friction is $\min(\mu_{\mathrm{wheel}}, 1.0) = \mu_{\mathrm{wheel}}$, giving per-environment control.
However, friction remains spatially uniform within each world: every wheel contact with the ground plane uses the same coefficient.
Real roads exhibit local friction variation (wet patches, oil spills, transitions between asphalt and concrete), and scenarios such as a slippery lane adjacent to a dry lane cannot be represented.
Modeling local friction would require either a segmented ground mesh with per-face materials or runtime material-property writes keyed to vehicle position, neither of which is straightforward under PhysX~5's current contact pipeline.

\paragraph{PhysX substep granularity.}
Physics runs at 120\,Hz with a decimation factor of 4, yielding 30\,Hz control.
The 120\,Hz rate is a compromise between simulation fidelity and throughput: higher rates (240\,Hz or 480\,Hz) would improve suspension stability and contact resolution but proportionally reduce training speed.
Our sysid calibration (\S\ref{sec:vehicle_sysid}) was performed at this rate, so the fitted parameters absorb any 120\,Hz-specific artifacts; re-calibration would be needed if the substep rate were changed.

\paragraph{Clone-in-Fabric disabled.}
Isaac Lab's \texttt{clone\_in\_fabric} optimization, which clones environment state entirely in the Fabric layer for maximum GPU throughput, is disabled in SceneFactory because it is incompatible with the post-clone road injection workflow described in \S\ref{sec:data_processing}.
Fabric assumes that all environments are structurally identical after cloning; SceneFactory's per-environment road prims violate this assumption, causing metadata reads to return stale or incorrect values.
Disabling Fabric increases CPU--GPU synchronization overhead but is necessary for correctness.

\paragraph{Constant-velocity TTC.}
Both the inter-vehicle and road-edge TTC computations assume constant velocity over the prediction horizon.
This systematically overestimates collision risk during deceleration (the vehicle is predicted to continue at its current speed toward the obstacle) and underestimates risk during acceleration.
A quadratic (constant-acceleration) TTC formulation would be more accurate, particularly in approach zones where vehicles are actively braking, but would increase the per-step computational cost of the $K_v{\times}9$ pairwise circle tests.

\section{Throughput Analysis Details}
\label{app:throughput}

\paragraph{Comparison with kinematic simulators.}
GPUDrive~\cite{kazemkhani_gpudrive_2025} (ICLR 2025) reports a peak of 2.3\,M agent-steps per second (ASPS) on an A100 GPU using a kinematic bicycle model.
When restricted to \emph{controlled} agents (those with non-trivial goals), the effective rate drops to $\sim$200\,K CASPS due to the high fraction of parked or near-goal vehicles in randomly sampled Waymo scenes (mean $\sim$10.8 controllable agents out of up to 128 slots).
Nocturne~\cite{vinitsky_nocturne_2022} (NeurIPS 2022), a CPU-based C++ simulator with a kinematic model, achieves $\sim$15\,K ASPS on 16 CPU cores.
We benchmarked Waymax~\cite{gulino_waymax_2023} on the same RTX PRO 6000 at 64 parallel scenes $\times$ 16 agent slots using \texttt{DeltaGlobal} kinematic dynamics, a scripted constant-speed actor, and a 350-point per-agent road-geometry observation; scenarios were drawn from a real WOMD training shard (455 unique scenes), cycling through four preloaded batches to avoid cache artifacts.
Under this configuration Waymax achieves $\sim$64\,M CASPS$^\dagger$.
The $\sim$3{,}300$\times$ gap reflects the difference in per-step cost: Waymax's kinematic step is a single trigonometric integration, whereas SceneFactory runs a 10-DOF PhysX articulated-body solver at 120\,Hz plus a live PPO policy forward pass.
These comparisons are not apples-to-apples; the relevant same-hardware, same-physics reference is the non-vectorized PhysX Baseline, where SceneFactory achieves a 127$\times$ speedup.

\paragraph{Comparison with MetaDrive (physics-based, measured on our hardware).}
MetaDrive~\cite{li2022metadrive} uses the Bullet rigid-body physics engine and, like the non-vectorized Baseline, issues actuation, observation, and reward computations through per-agent Python loops.
We run \texttt{MultiAgentRoundaboutEnv} at agent counts of 1, 4, 8, 16, 32, and 40 (the spawn-geometry ceiling for this map), using random actions to isolate loop overhead.
CASPS is measured over 10 steady-state windows of 500 steps each after a 200-step warmup.
MetaDrive peaks at $\sim$1{,}390 CASPS with 8 agents and declines to 1{,}050 CASPS at 40 agents, confirming that per-agent Python loops, not the physics solver, are the throughput bottleneck regardless of whether the backend is PhysX~5 or Bullet.
MetaDrive has no multi-world parallelism axis; SceneFactory's \texttt{num\_worlds} dimension is architecturally absent.
At the smallest SceneFactory configuration (32 worlds, 512 agent slots) SceneFactory achieves 3{,}870 CASPS, a 2.8$\times$ advantage over MetaDrive's per-scene peak, growing to 13.9$\times$ at 256 worlds.

\begin{table}[h]
\centering
\caption{Cross-simulator throughput comparison.
All kinematic simulators use simplified vehicle dynamics (no rigid-body contact, no tire model).
ASPS = agent steps per second (all agents, including parked); CASPS = controlled agent steps per second (policy-active agents only).
GPUDrive and Nocturne numbers are from~\cite{kazemkhani_gpudrive_2025}; Waymax, MetaDrive and SceneFactory are measured on our RTX~PRO~6000.
$^\dagger$Waymax uses a scripted constant-speed actor with kinematic dynamics; SceneFactory is measured during live PPO training with full PhysX articulated-body dynamics.
MetaDrive's agent count is bounded by spawn-road geometry ($\leq 40$) with no multi-world axis.}
\label{tab:cross_simulator}
\small
\begin{tabular}{llllr}
\toprule
Simulator & Physics model & GPU & Metric & Throughput \\
\midrule
GPUDrive~\cite{kazemkhani_gpudrive_2025} & Kinematic bicycle & A100 80\,GB & ASPS & 2{,}300{,}000 \\
GPUDrive (training) & Kinematic bicycle & A100 80\,GB & CASPS & 200K--500K \\
Nocturne~\cite{vinitsky_nocturne_2022} & Kinematic & CPU 16 cores & ASPS & 15{,}000 \\
Waymax~\cite{gulino_waymax_2023} & Kinematic bicycle & RTX PRO 6000 & CASPS$^\dagger$ & 64{,}338{,}897 \\
\midrule
MetaDrive~\cite{li2022metadrive} (8 agents) & Bullet rigid body & RTX PRO 6000 & CASPS & 1{,}390 \\
MetaDrive~\cite{li2022metadrive} (40 agents) & Bullet rigid body & RTX PRO 6000 & CASPS & 1{,}050 \\
\midrule
\textbf{SceneFactory (ours)} & \textbf{10-DOF rigid body} & \textbf{RTX PRO 6000} & \textbf{CASPS} & \textbf{19{,}250} \\
\bottomrule
\end{tabular}
\end{table}

\paragraph{Timing breakdown.}
Table~\ref{tab:app_timing_breakdown} reports the per-step timing breakdown for both the Baseline and SceneFactory pipelines at 64 worlds (1{,}024 agent slots, all active), from the Experiment~A scaling benchmark on scene \texttt{000077}.

On the Baseline side, the PhysX rigid-body solver accounts for roughly half of wall-clock time (2{,}934\,ms, 51.6\%); this cost is irreducible and shared by both pipelines.
The observation builder consumes another 1{,}456\,ms (25.6\%), because it iterates over each of the 1{,}024 active agents in a Python loop, issuing per-prim USD state queries and computing ego-frame transforms with scalar arithmetic.
Geometric lane features, TTC computation, and action application add a further 22\% through similar per-agent loops.
The total per-step cost is 5{,}683\,ms.

On the SceneFactory side, the same 64-world configuration completes an environment step in 135\,ms, a 42$\times$ reduction.
Physics remains the largest single cost (53.8\,ms), but the GPU-batched solver is 55$\times$ faster than the Baseline's single-scene PhysX dispatch.
Observations, rewards, dones, and actions collectively take only 78\,ms because they execute as batched \texttt{torch} tensor operations with no per-agent Python loop.

\begin{table}[t]
\centering
\caption{Per-step timing breakdown at 64 worlds (1{,}024 agent slots, all active) from the Experiment~A benchmark on scene \texttt{000077}.
Baseline sub-phases are grouped into the matching SceneFactory category.
SceneFactory sub-phases may overlap on the GPU, so they sum to more than the measured wall-clock total; the \textbf{Total} row reports the true measured time.}
\label{tab:app_timing_breakdown}
\small
\begin{tabular}{lrrr}
\toprule
Phase & Baseline (ms) & SceneFactory (ms) & Speedup \\
\midrule
Physics solver         & 2{,}934 &  53.8 & 55$\times$ \\
Observation builder    & 1{,}772 &  23.3 & 76$\times$ \\
Reward \& termination  &     629 &  29.9 & 21$\times$ \\
Action application     &     299 &  24.9 & 12$\times$ \\
Reset / overhead       &      49 &  32.4 &  1.5$\times$ \\
\midrule
\textbf{Total (wall-clock)} & \textbf{5{,}683} & \textbf{135} & \textbf{42$\times$} \\
\bottomrule
\end{tabular}
\\[4pt]
{\footnotesize
\emph{Baseline grouping:}
Observation builder = obs.\ construction (1{,}456\,ms) + geometric lane features (316\,ms);
Reward \& termination = road-edge TTC (236) + vehicle TTC (191) + TTC state queries (136) + contact scan (66).
\emph{SceneFactory grouping:}
Reward \& termination = reward computation (18.4) + done/termination (11.5);
Physics solver = write (33.0) + simulate (15.9) + read-back (4.9).}
\end{table}

\paragraph{Why the Baseline is slow.}
SceneFactory eliminates Baseline bottlenecks through three architectural changes:
\begin{enumerate}[leftmargin=*, itemsep=2pt]
  \item \textbf{Batched joint-state tensors.}
        The custom URDF articulated vehicle (\S\ref{sec:vehicle_sysid}) exposes positions, velocities, and orientations as GPU-resident tensors of shape $(\texttt{num\_envs}, \texttt{num\_agents}, d)$ via Isaac Lab's \texttt{Articulation} API.
        Observation construction reduces to a handful of batched \texttt{torch} operations with no Python per-agent loop.
  \item \textbf{Cloned PhysX scenes.}
        Isaac Lab's environment cloning creates $N$ independent PhysX solver islands, enabling the GPU broad-phase and narrow-phase to dispatch them in parallel.
        The Baseline places all worlds in a single PhysX scene, limiting solver parallelism.
  \item \textbf{GPU-resident training loop.}
        RSL-RL reads observations, rewards, and dones as CUDA tensors directly from the environment; no CPU round-trip occurs during training.
        The Baseline converts observations to NumPy for Stable-Baselines3's rollout buffer, incurring $O(\text{agents} \times \text{obs\_dim})$ bytes of device-to-host transfer every step.
\end{enumerate}

\section{Friction Awareness Ablation}
\label{app:friction_ablation}

We compare two policies that differ only in their friction training regime.
\textbf{v7 (friction-aware)} was trained with per-world weather variation (10\% of worlds assigned a wet surface drawn uniformly from SMA, AC, and OGFC at varying water film thicknesses); the policy receives a 4-element weather context vector $[f_\text{dry}, f_\text{SMA}, f_\text{AC}, f_\text{OGFC}]$ at each step (checkpoint: \texttt{model\_600}, $\sim$314M steps).
\textbf{v8 (friction-blind)} was trained exclusively on dry asphalt with the weather context fixed at $[0,1,0,0]$; only the underlying PhysX friction changes at evaluation time (checkpoint: \texttt{model\_300}, $\sim$314M steps).
Both use the same vehicle (sysid4), reward (v8-fastgoal), 256-world random scene pool, and PPO hyperparameters.

\paragraph{Test configurations.}
We evaluate on 64 held-out test scenes under three controlled surface assignments:
(1)~dry asphalt (AC, $\mu{=}1.1$, original dataset OD pairs);
(2)~moderate wet (AC 0.5\,mm water film, $\mu{=}0.859$, original OD);
(3)~hard rain (SMA 2\,mm, $\mu\approx10^{-3}$, original OD) — included as a degenerate-traction reference.
All evaluations use invincible mode, 50\,s episodes, and 3.0\,m goal threshold.

\begin{table}[h]
\centering
\caption{Friction awareness ablation: v7 (friction-aware, \texttt{model\_600}) vs.\ v8 (friction-blind, \texttt{model\_600}).
All evaluations use invincible mode, 50\,s episodes, and 3.0\,m goal threshold.
DRAC = deceleration rate to avoid crash (m/s$^2$).}
\label{tab:friction_ablation}
\small
\begin{tabular}{llccc}
\toprule
Surface & Model & SR $\uparrow$ & CR $\downarrow$ & Mean max DRAC $\downarrow$ \\
\midrule
\multirow{2}{*}{\makecell[l]{Dry\\(AC, $\mu{=}1.1$)}}
  & v7 & 77.2\% & 5.1\% & 37.1 \\
& v8  & 77.9\% & 6.3\% & 50.2 \\
\midrule
\multirow{2}{*}{\makecell[l]{Moderate wet\\(AC 0.5\,mm, $\mu{=}0.859$)}}
  & v7  & 77.9\% & 5.9\% & \textbf{27.8} \\
& v8   & 78.0\% & 6.3\% & \textbf{58.7} \\
\midrule
\multirow{2}{*}{\makecell[l]{Hard rain\\(SMA 2\,mm, $\mu{\approx}10^{-3}$)}}
  & v7  & 8.2\%  & 0.6\% & 0.93 \\
& v8  & 8.0\%  & 1.0\% & 0.99 \\
\bottomrule
\end{tabular}
\end{table}

\paragraph{Discussion.}
On both dry and moderate-wet surfaces, goal-reach SR is statistically indistinguishable ($\Delta\text{SR}\leq0.8$\,pp), a null result for friction-aware conditioning on navigation success.
We attribute this to a \emph{credit-assignment cost}: the weather-context vector is seen in only 10\% of v7's training worlds, providing insufficient gradient signal within $\sim$314M steps.

The moderate-wet DRAC result reveals a subtler behavioral effect: v7's mean max DRAC drops from 37.1 (dry) to 27.8 (wet), while v8's rises from 50.2 to 58.7 — a 2.1$\times$ gap absent on dry road.
This is consistent with v7 learning gentler braking under reduced traction, even without a corresponding SR improvement on short routes.

The hard-rain (SMA 2\,mm, $\mu\approx10^{-3}$) condition is a degenerate-traction regime: both policies collapse to $\sim$8\% SR because the friction floor ($10^{-3}$) makes cornering and braking nearly impossible regardless of policy.
The near-identical collapse confirms that physics dominates over policy in this regime, and validates that the friction module correctly imposes the intended physical constraint.
Increasing wet-world training exposure and probing intermediate film thicknesses are the most promising directions for eliciting stronger friction-conditioning benefits.

\section{PPO Hyperparameters}
\label{app:ppo_hparams}

Table~\ref{tab:app_ppo_hparams} lists the full PPO configuration used in all experiments.
The learning rate follows an adaptive schedule that halves or doubles the rate to maintain the per-update KL divergence near a target of 0.01.

\begin{table}[h]
\centering
\caption{PPO hyperparameters.}
\label{tab:app_ppo_hparams}
\small
\begin{tabular}{ll}
\toprule
Parameter & Value \\
\midrule
Rollout horizon           & 128 steps \\
Mini-batches              & 16 \\
Learning epochs           & 5 \\
Learning rate (initial)   & $2.09 \times 10^{-4}$ \\
LR schedule               & adaptive (KL target 0.01) \\
Discount $\gamma$         & 0.99 \\
GAE $\lambda$             & 0.98 \\
Clip $\epsilon$           & 0.15 \\
Entropy coefficient       & $3 \times 10^{-5}$ \\
Value loss coefficient    & 1.0 \\
Max gradient norm         & 1.0 \\
\bottomrule
\end{tabular}
\end{table}


\end{document}